\documentclass[]{mn2e}

\usepackage{amssymb}
\usepackage{graphicx}
\usepackage{amsmath}

\usepackage{graphics}
\usepackage{color}
\title[Effects of disorder on grain spectra]{Effects of structural and chemical disorders on the vis/UV spectra of carbonaceous interstellar grains}
\author[R.J. Papoular, S. Yuan, R. Roldan et al.]
{Robert~J.~Papoular,$^{1}$\thanks{E-mail: Robert.Papoular@cea.fr}
Shengjun~Yuan,$^{2}$\thanks{E-mail: s.yuan@science.ru.nl} Rafael~Rold\'{a}n,$^{3}$
\thanks{E-mail: rroldan@icmm.csic.es}
 Mikhail~I.~Katsnelson$^{2}$\thanks{E-mail: M.Katsnelson@science.ru.nl}
\newauthor 
and Renaud~Papoular$^{4}$\thanks{E-mail: papoular@wanadoo.fr}\\
$^{1}$IRAMIS, Laboratoire Leon Brillouin, CEA Saclay, 91191 Gif-s-Yvette, France\\
$^{2}$Institute for Molecules and Materials, Radboud University of Nijmegen, 6525AJ Nijmegen, The Netherlands\\
$^{3}$Instituto de Ciencia de Materiales de Madrid, CSIC, Cantoblanco E28049 Madrid, Spain\\
$^{4}$Service d'Astrophysique and Service de Chimie Moleculaire,
 CEA Saclay, 91191 Gif-s-Yvette, France}

\begin{document}
   \maketitle
\label{firstpage}

\begin{abstract}

The recent spectacular progress in the experimental and theoretical understanding of graphene, the basic constituent of graphite, is applied here to compute, from first principles, the UV extinction of nano-particles made of stacks of graphene layers. The theory also covers cases where graphene is affected by structural, chemical or orientation disorder, each disorder type being quantitatively defined by a single parameter. The extinction bumps carried by such model materials are found to have positions and widths falling in the same range as the known astronomical 2175~\AA~ features: as the disorder parameter increases,
the bump width increases from 0.85 to 2.5 $\mu$m$^{-1}$, while its peak position shifts from 4.65 to 4.75 $\mu$m$^{-1}$. Moderate degrees of disorder are enough to cover the range of widths of the vast majority of observed bumps (0.75 to 1.3 $\mu$m$^{-1}$). Higher degrees account for outliers, also observed in the sky.

The introduction of structural or chemical disorder amounts to changing the initial $sp^{2}$ bondings into $sp^{3}$ or $sp^{1}$, so the optical properties of the model material become similar to those of the more or less amorphous carbon-rich materials studied in the laboratory: a-C, a-C:H, HAC, ACH, coals etc. The present treatment thus bridges gaps between physically different model materials.
\end{abstract}



\section{Introduction}

The ISEC (InterStellar Extinction Curve) is dominated, in the vis/UV, by a strong and ubiquitous feature, peaking at 5.7 eV (or 4.6 $\mu$m$^{-1}$). The discoverers of this feature suggested that it might be carried by small crystalline graphite particles (Stecher and Donn 1965). Pure bulk crystalline graphite in the laboratory displays an extinction feature near 4.25 eV, but, in fine powder form, its extinction peaks at 5.7 eV and was assigned to a ``surface plasmon mode'', or Fr\"ohlich resonance, due to the $\pi$ electron of graphite (Bohren and Huffman, 1983); it is also called ``bump'' in astronomer parlance. Since then, however, together with developments in this direction, several other candidate carriers have been predicated, ranging from molecular to solid, more or less hydrogenated and more or less graphitic. A necessarily very short and partly subjective sample of the latest of these models includes mixtures of PAHs (Cecchi-Pestellini et al. 2008); Bucky Onions (fullerenes); fullerenes/buckyonions models (e.g. De Heer and Ugarte 1993, Wada and Tokunaga 2006, Chhowalla et al. 2003, Ruiz et al. 2005, Li et al. 2008); mixtures of stacks of PAH molecules and HAC/a-C:H (hydrogenated amorphous carbon (Duley and Seahra 1999); mixtures of ellipsoidal amorphous carbon grains (Mennella et al. 1998); agglomerated HAC grains (Schnaiter et al. 1998 ), and PCG (polycrystalline graphite, Papoular and Papoular 2009). The common purpose is to  satisfy both tight constraints imposed by astronomical observations: relatively large variations of width (roughly 0.7 to 1.3 $\mu$m$^{-1}$) while the peak position remains fixed at 4.6 $\mu$m$^{-1}$ to $\sim$2 $\%$ (see Fitzpatrick and Massa 1986, 2007). Figure 16 of Fitzpatrick and Massa \cite{fm07} plots all observed features in a graph of the peak wavenumber, $\nu_{0}$, against the FWHM (Full Width Half-Maximum) width, $\gamma$. It shows that the vast majority of representative points lie within a cluster centered at ($\sim0.9 \,\mu$m$^{-1}$, 4.59\, $\mu$m$^{-1}$), with dimensions ($\sim0.3 \,\mu$m$^{-1}$, 0.1\, $\mu$m$^{-1}$). However, many outliers are also observed, which seem to extend this cloud towards larger widths, with rapidly decreasing density, as confirmed by the histograms of fig. 17 of the same reference (see also, for instance, Cardelli and Savage 1988).

A large number of carbonaceous materials are found to exhibit the $\pi$ and $\sigma$ UV resonances of graphite with various relative intensities, even if highly loaded with hydrogen. In general, the former decrease considerably as structural disorder and chemical impurities increase. This is illustrated by artificially produced a-C:H (see for instance Fink et al. 1984) as well as natural coals extracted from mines of decreasing depth, i.e. of decreasing degree of graphitization. Papoular et al.\cite{pap93} and Papoular et al. \cite{pap95} showed that the $\pi$ feature is exhibited not only by graphite, but also by coals with decreasing degree of graphitization, until it is completely damped by excessive H content and $sp^{3}$ bondings. The inverse evolution can be induced, starting from the less graphitized coal or kerogen, by thermal annealing and/or high energy irradiation. Mennella et al. \cite{men95} and Mennella et al. \cite{men98}, studying artificially produced carbonaceous materials, also observed and used this general property of carbon based materials.

For all types of materials, it was found that the $\sigma$ feature can still be observed when the $\pi$ feature has long been drowned. This is to be expected since the former is associated with $pp\sigma$ types of bonding (in which principal axes of coupled $p$ orbitals lie along the same line; see Castro Neto et. al. \cite{cas}), which survive the transition to $sp^{3}$ hybridization that dominates structurally disordered materials. Robertson \cite{rob}, in his study of HAC, compared extensively the optical and other properties of graphite with those of disordered carbons (HAC, glassy carbon, a-C, a-C:H, etc.). It is striking to see the kind of continuity linking these various carbon-rich materials and their transformation from one into another through irradiation and/or thermal annealing. Several attempts at modelling these disordered materials were made, with a view to understanding the physics underlying their spectral properties. For instance, Stenhouse and Grout \cite{ste} proposed the ``domain model'', which pictures a-C as microcrystallites of graphite inter-linked by an $sp^{3}$-bonded random network. In the random network model of Beeman et. al. \cite{bee}, the structure consists of $sp^{2}$ and $sp^{3}$ sites in different proportions (see, also Robertson 1991). Mennella et al. \cite{men98} used continuous distributions of ellipsoids.

Still another kind of disorder is obtained with pure PCG: this is orientation disorder. Papoular and Papoular \cite{pap09} showed that industry-made pellets of compressed powder of pure graphite (which can be considered as artificially made disordered graphite)  exhibit a feature at 2200~ \AA (4.52\, $\mu$m$^{-1}$), 1.3\, $\mu$m$^{-1}$ in width. This is one of the few largest widths observed in the sky by Fitzpatrick and Massa \cite{fm86}.

\begin{figure*}
\mbox{
\includegraphics[width=7cm]{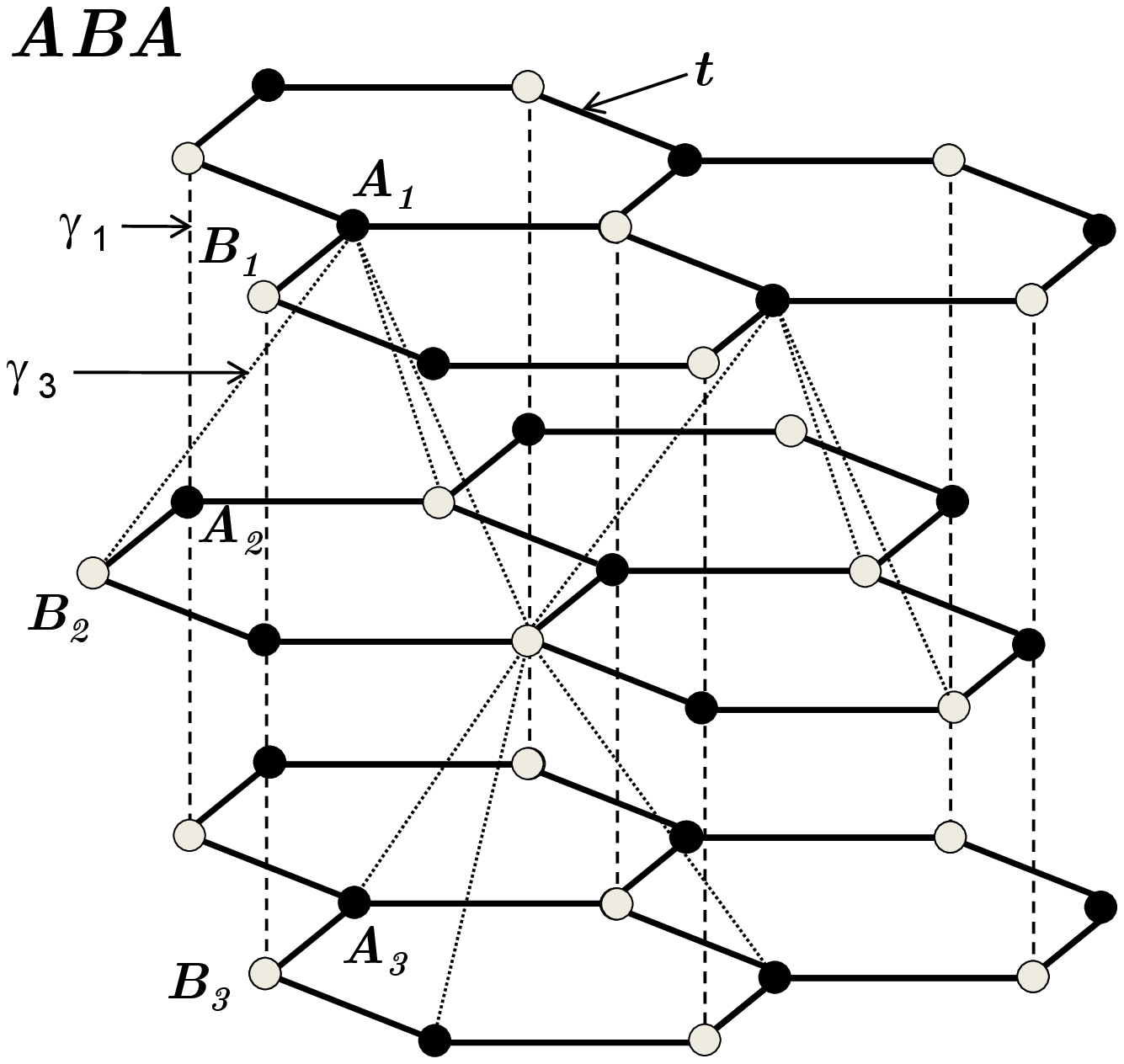}
\includegraphics[width=7cm]{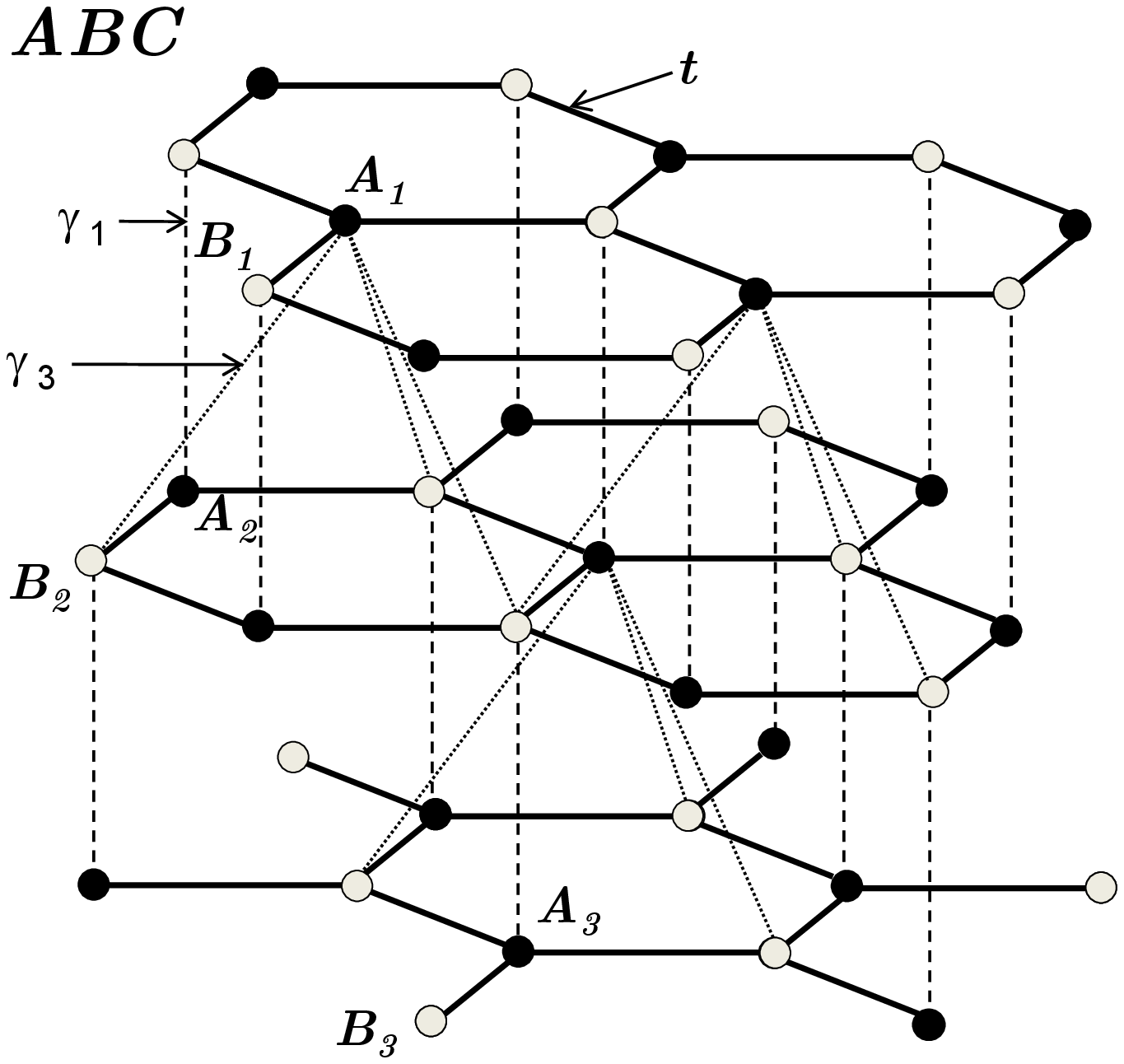}
}
\caption[]{Atomic structure of the two different stacking sequences of the graphene layers considered in this work: ABA- (left) and ABC-stacked (right) MLG. The nearest neighbor
intra-layer $t$ and the inter-layer $\protect\gamma _{1}$ and $\protect\gamma
_{3}$ hopping amplitudes are schematically shown in the figure.}
\label{Fig:Stacking}
\end{figure*}

It appears, therefore, that a vast gamut of carbonaceous materials display the UV bump, and that the latter's profile depends very much on the amount of structural and chemical disorder in the carrier. The problem, then, is to model this disorder as generally as possible, based on the best available solid state theory and in such a way as to allow a straightforward derivation of the optical properties as a function of type and degree of disorder, at least over the whole range of the $\pi$ resonance, with the minimum possible $ad \,hoc$ assumptions and adjustable parameters. This should also help understanding why most of the UV bump widths flock within a small interval and why they are limited from below.

For this purpose, we start, here, with the carrier model consisting of stacks of graphitic layers in the form of randomly oriented small spherical grains, as studied by Draine and Lee \cite{dl}. These authors invoke dielectric functions derived from measurements on bulk graphite. In this paper, we derive these functions from an accurate tight-binding description, based on the dramatic progress made in recent years in the theoretical and experimental study of graphene in single and multiple layers (SLG, MLG), the constitutive elements of graphite. Three of us (S.Y., R.R. and M.K.) have been engaged in this research for some time and have recently completed a theoretical model of the electronic and optical properties of SLG and MLG from the infrared to the UV, encompassing the $\pi$ resonance of $sp^{2}$ bondings (Yuan et al. 2011a, 2011b). This model has been successfully applied to account for different experimental measurements, as the $\pi$-plasmon resonance observed by electron energy loss spectroscopy (EELS) (Eberlein et al. 2008 and Gass et al. 2008), the effect of disorder on the optical conductivity spectrum of doped graphane (Li et al. 2008), or the screening properties and the polarization function obtained from inelastic X-ray scattering (Reed et al. 2010). The method allows for the consideration of different types of non-correlated and correlated disorder, such as random or Gaussian potentials, random or Gaussian nearest-neighbor hopping parameters, randomly distributed vacancies or their clusters, and randomly adsorbed hydrogen atoms or their clusters. We notice that, for the same amount of disorder, the electronic properties of SLG and MLG are less affected by the Gaussian correlated disorder, as the symmetry of the clean sample is less broken as compared to the non-correlated disorder (Yuan et al. 2011b). In fact, completely randomly distributed (non-correlated) impurities lead to the largest effect on the optical spectrum. Therefore, we consider here four different sources of non-correlated disorder, each of them controlled by a single parameter: 1) random on-site potential, corresponding to a local shift of chemical potential (inhomogeneous distribution of electron and hole puddles), 2) random hopping parameter between two nearest neighbors, corresponding to random variations of inter-atomic distances and directions, 3) randomly distributed vacancies, corresponding to atomic defects with missing carbon atoms (presence of $ sp^{1}$ bonds in the $sp^{2}$ networks), and 4) randomly distributed hydrogen-like resonant impurities (including organic molecules such as CH$_{3}$, C$_{2}$H$ _{5}$, CH$_{2}$OH, as well as H and OH groups).

We consider mostly $sp^{2}$-conjugated carbon in stacks of finite graphene sheets, arranged in the graphitic stacking schemes ABA (Bernal type), or  ABC (rhombohedral), which are sketched in Fig. \ref{Fig:Stacking}, and perturbed by the different kinds of disorder introduced above. While our results clarify a number of issues concerning graphene and graphite properly, here we are mainly interested in the impact of impurities and disorder on the position and width of the $\pi$ surface plasmon (UV bump). As is well known (Katsnelson 2012), the band structure of graphene presents saddle points (Van Hove singularities), with the corresponding logarithmic divergencies in the density of states (DOS), at energies $E= \pm t$ (in the nearest neighbor approximation), where $t$ is the nearest neighbor hopping parameter.  The $\pi$-plasmon mode is associated to electron-hole transitions between the Van Hove singularities of the valence ($\pi$) and the conduction ($\pi^*$) bands. As the disorder parameter increases, it is found that the slope of the blue wing of the $\pi$ resonance decreases in such a way as to increase the width of the plasmon feature, leaving its peak position nearly unchanged, as required by astronomical observations. This trend is common to all types of disorder considered, but is more or less pronounced.
 Our results show that random on-site potential disorder is the most effective, but, for all types, at sufficiently high degrees of disorder, the $\pi$ resonance is completely damped, indicating that the range of our computations covers the most extreme situations, from perfect order to very high disorder observed on laboratory materials. The UV bump survives considerable disorder, at the price of broadening and frequency shift.

This paper is organized as follows. Section 2 lays down the basic principles, assumptions and approximations underlying our calculations, and discusses various types of disorder (imperfections) considered here. Section 3 displays some of our results in the form of the electrical conductance of graphene as a function of normalized wave number, for different types and degrees of disorder. Section 4 describes the procedure which delivers the corresponding dielectric function and, hence, the extinction cross-section. Section 5 displays a number of computed UV bump profiles and highlights their changes as a function of disorder. In Sec. 6, our computed results are used to derive the extinction bumps of free-flying graphitic bricks, and these are compared with those of a few disordered laboratory model carrier candidates. Finally, in Sec. 7, all these cases are confronted with available astronomical observations. A discussion of these results and their possible improvements and applications concludes this work .

\section{Theoretical procedure}

 We consider the effect of different sources of disorder on SLG and MLG by direct numerical simulations of electrons on a honeycomb lattice in the framework of the tight-binding approximation. This allows us, by means of the  time-evolution method, to obtain the DOS of large samples containing millions of atoms. The time-evolution method is based on the numerical solution of the time-dependent Schr\"odinger equation (TDSE) with additional averaging over random superposition of basis states (Hams et al. 2000, Yuan et al. 2010). The electronic single particle dispersion in SLG is properly described by the hopping of electrons between nearest neighbor carbon sites arranged in a honeycomb lattice with two atoms per unit cell. For MLG, additional inter-layer hopping processes must be included in the kinetic Hamiltonian. A sketch of a MLG with the different hopping parameters used in the calculation is shown in Fig. \ref{Fig:Stacking}.

The tight-binding Hamiltonian of a MLG is given by
\begin{equation}
H=\sum_{l=1}^{N_{layer}}H_{l}+\sum_{l=1}^{N_{layer}-1}H_{l}^{\prime } + H_{imp},
\label{Hamiltonian}
\end{equation}%
where $H_{l}$ is the Hamiltonian of the $l$'th layer of graphene,%
\begin{equation}
H_{l}=-\sum_{\langle i,j\rangle}(t_{l,ij}a_{l,i}^{\dagger }b_{l,j}+\mathrm{h.c}%
)+\sum_{i}v_{l,i}c_{l,i}^{\dagger }c_{l,i},  \label{Eq:H_SLG}
\end{equation}%
where $a_{l,i}^{\dagger }$ ($b_{l,i}$) creates (annihilates) an electron on
sublattice A (B) of the $l$'th layer, and $t_{l,ij}$ is the nearest neighbor
hopping parameter if $i,j$ are nearest neighbors and zero otherwise. It is known 
that the nearest-neighbor approximation is quite accurate to describe electronic 
structure of singla-layer graphene (Castro-Neto et al. 2009, Katsnelson 2012). The second term of $H_l$ 
accounts for the effect of an on-site potential $v_{l,i}$, where $n_{l,i}=c^{\dagger}_{l, i}c_{l, i}$ is
the occupation number operator and $c=a,b$. In the second term of the Hamiltonian Eq. (%
\ref{Hamiltonian}), $H_{l}^{\prime }$ describes the hopping of electrons
between layers $l$ and $l+1$. We consider the two more stable forms of
stacking sequence in bulk graphite, namely ABA (Bernal) and ABC
(rhombohedral) stacking, as shown in Fig. \ref{Fig:Stacking}. For a MLG with an ABA stacking, $H_{l}^{\prime }$ is given by%
\begin{equation}
H_{l}^{\prime }=-\gamma _{1}\sum_{j}\left[ a_{l,j}^{\dagger }b_{l+1,j}+%
\mathrm{h.c.}\right] -\gamma _{3}\sum_{j,j^{\prime }}\left[ b_{l,j}^{\dagger
}a_{l+1,j^{\prime }}+\mathrm{h.c.}\right] ,  \label{Eq:H-interlayer}
\end{equation}%
where the inter-layer hopping terms $\gamma_1$ and $\gamma_3$ are shown in
Fig. \ref{Fig:Stacking}. Thus, all the even layers ($l+1$) are rotated with
respect to the odd layers ($l$) by $+60^{\circ }$. The difference between
ABA and ABC stacking is that, the third layer(s) is rotated with respect to
the second layer by $-60^{\circ }$ (then it will be exactly under the first
layer) in ABA stacking, but by $+60^{\circ }$ in ABC stacking (Yuan et al. 2011a). The spin degree of freedom contributes only
through a degeneracy factor and is omitted for simplicity in Eq.~(\ref{Hamiltonian}).

The term $H_{imp}$ describes the hydrogen-like resonant impurities:
\begin{equation}
H_{imp}=\varepsilon _{d}\sum_{i}d_{i}^{\dagger }d_{i}+V\sum_{i}\left(d_{i}^{\dagger }c_{i}+\mathrm{h.c}\right) ,  \label{Eq:Himp}
\end{equation}
where $\varepsilon _{d}$ is the on-site potential on the impurity and $V$ is the hopping between carbon and absorbed atom. The band parameters $V\approx 2t$ and $\epsilon _{d}\approx -t/16$ are obtained from the \textit{ab initio} density functional theory (DFT) calculations (Wehling et al. 2010). 
Following Refs. (Wehling et al. 2010, Yuan et al. 2010, 2011b), we call these impurities as adsorbate hydrogen atoms but actually, the parameters for organic groups are almost
the same (Wehling et al. 2010). The band parameters for OH-group and fluorine are similar to those for hydrogen except that the on-site potential on impurities is $\epsilon _{d}\approx -t$ (Yuan et al. 2012). A vacancy is an atomic defect with one carbon atom missing, and can also be regarded as an atom (lattice point) with on-site energy $v_{i}\rightarrow \infty $ or with its hopping parameters to other sites being zero.

The numerical calculations of the optical conductivity and DOS are performed based on the numerical solution of the TDSE for the non-interacting electrons. The frequency-dependent
optical conductivity follows (Ishihara 1971, Yuan et al. 2010)
\begin{eqnarray}
\sigma _{\alpha \beta }\left( \omega \right) &=&\lim_{\varepsilon
\rightarrow 0^{+}}\frac{e^{-\beta \omega }-1}{\omega \Omega }
\int_{0}^{\infty }e^{-\varepsilon t}\sin \omega t  \nonumber
\\
&&\times 2~{\rm Im} \left\langle \varphi |f\left( H\right) J_{\alpha }\left(
t\right) \left[ 1-f\left( H\right) \right] J_{\beta }|\varphi \right\rangle
dt,  \nonumber \\
&&\label{Eq:OpCond}
\end{eqnarray}
(we take $\hbar =1$) where $\beta =1/k_{B}T$ is the inverse temperature, $\Omega $ is the sample area, $f\left( H\right) =1/\left[ e^{\beta \left(H-\mu \right) }+1\right] $ is the Fermi-Dirac distribution operator, $J_{\alpha }\left( t\right) =e^{iHt}J_{\alpha }e^{-iHt}$ is the time-dependent current operator in the $\alpha $ ($=x$ or $y$) direction, and $\left\vert \varphi \right\rangle $ is a random superposition of all the basis states in the real space, i.e., (Hams and De Raedt 2000, Yuan et al. 2010)
\begin{equation}
\left\vert \varphi \right\rangle =\sum_{i}a_{i}c_{i}^{\dagger }\left\vert 0\right\rangle ,  \label{Eq:phi0}
\end{equation}
where $a_{i}$ are random complex numbers normalized as $\sum_{i}\left\vert a_{i}\right\vert ^{2}=1$, and $\left\vert 0\right\rangle $ is the vacuum state. The time evolution operator $e^{-iHt}$ and the Fermi-Dirac
distribution operator $f\left( H\right) $ are obtained by the standard Chebyshev polynomial representation (Yuan et al. 2010).

\begin{figure*}
\mbox{
\includegraphics[width=9cm]{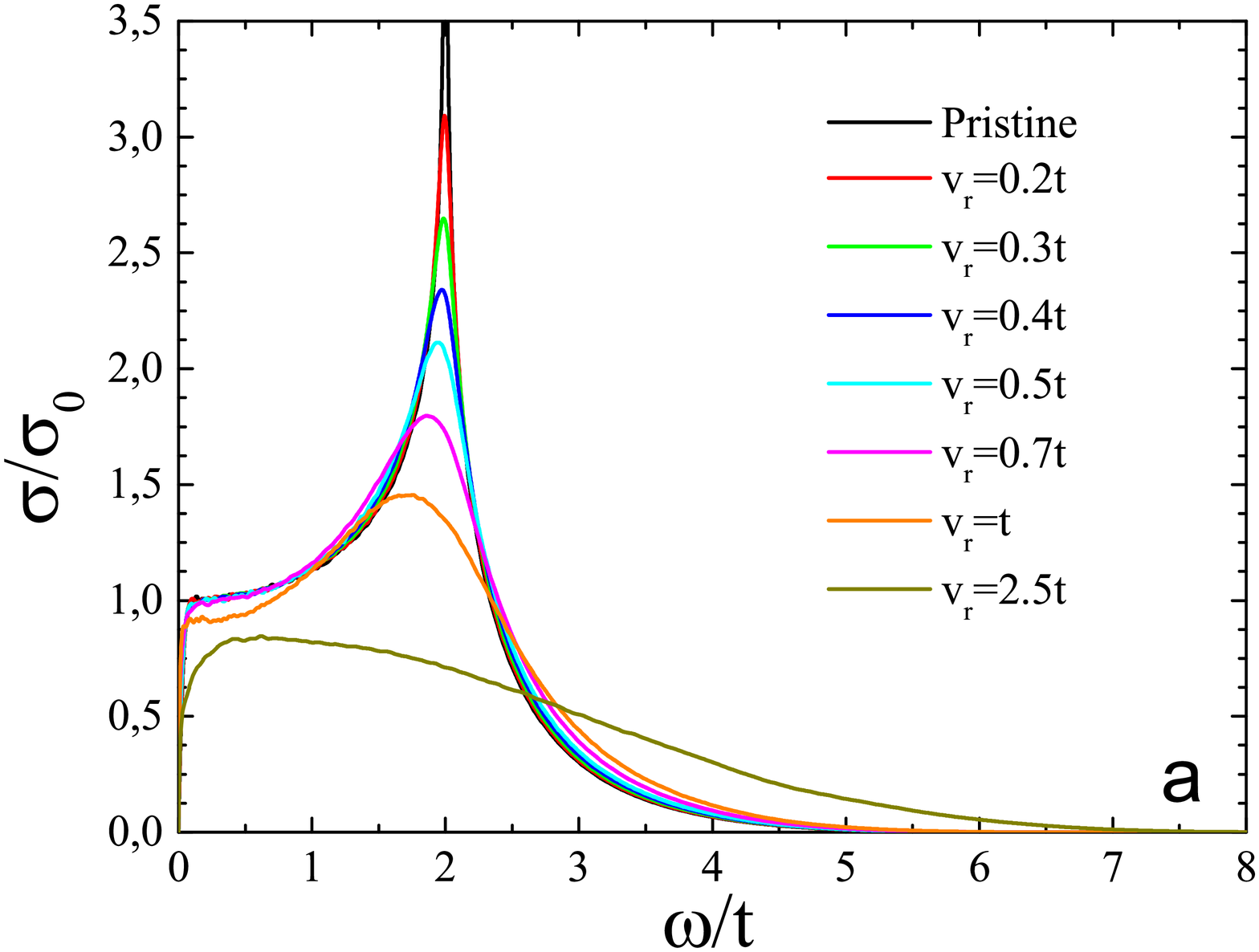}
\includegraphics[width=9cm]{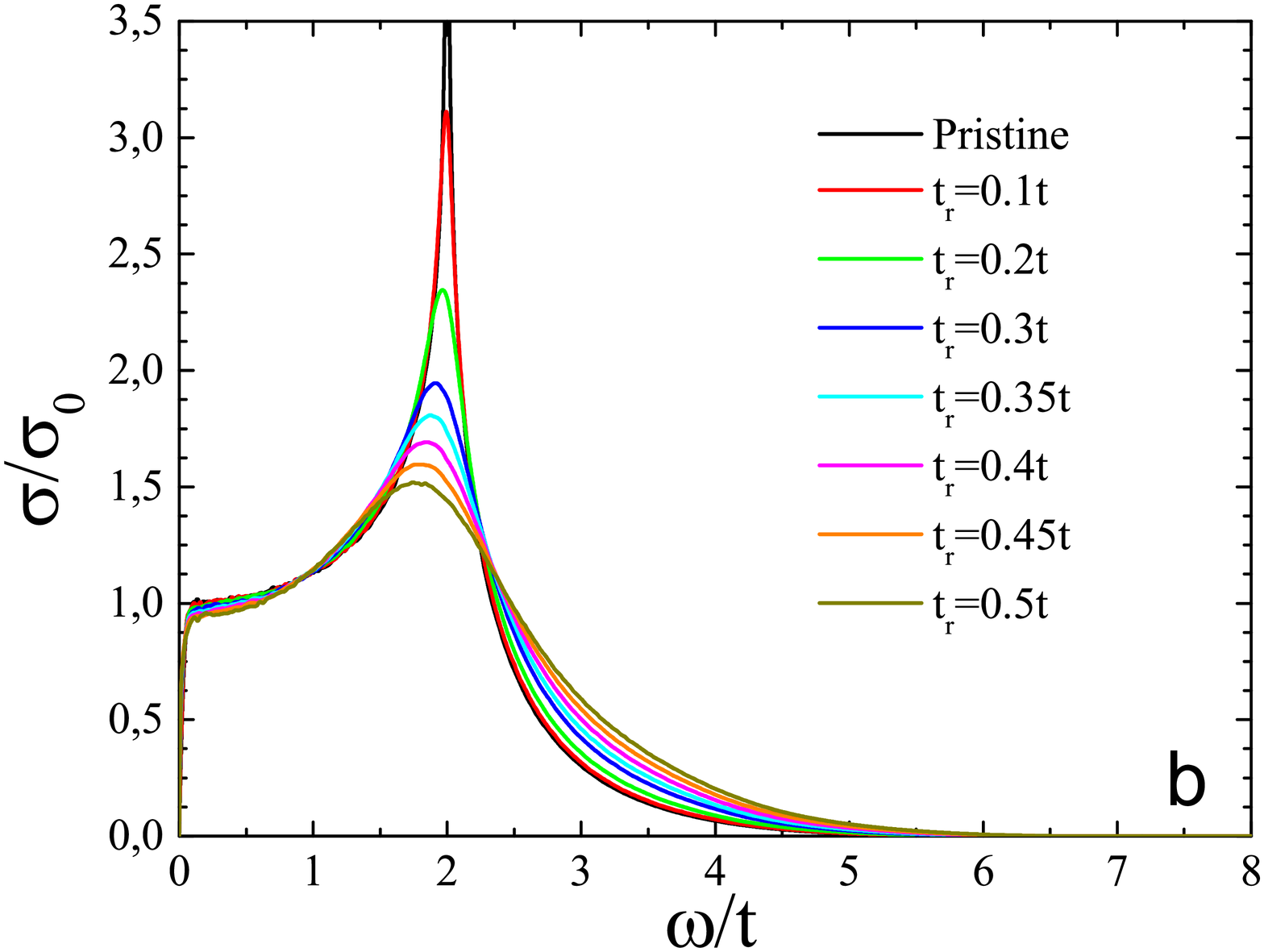}
}
\mbox{
\includegraphics[width=9cm]{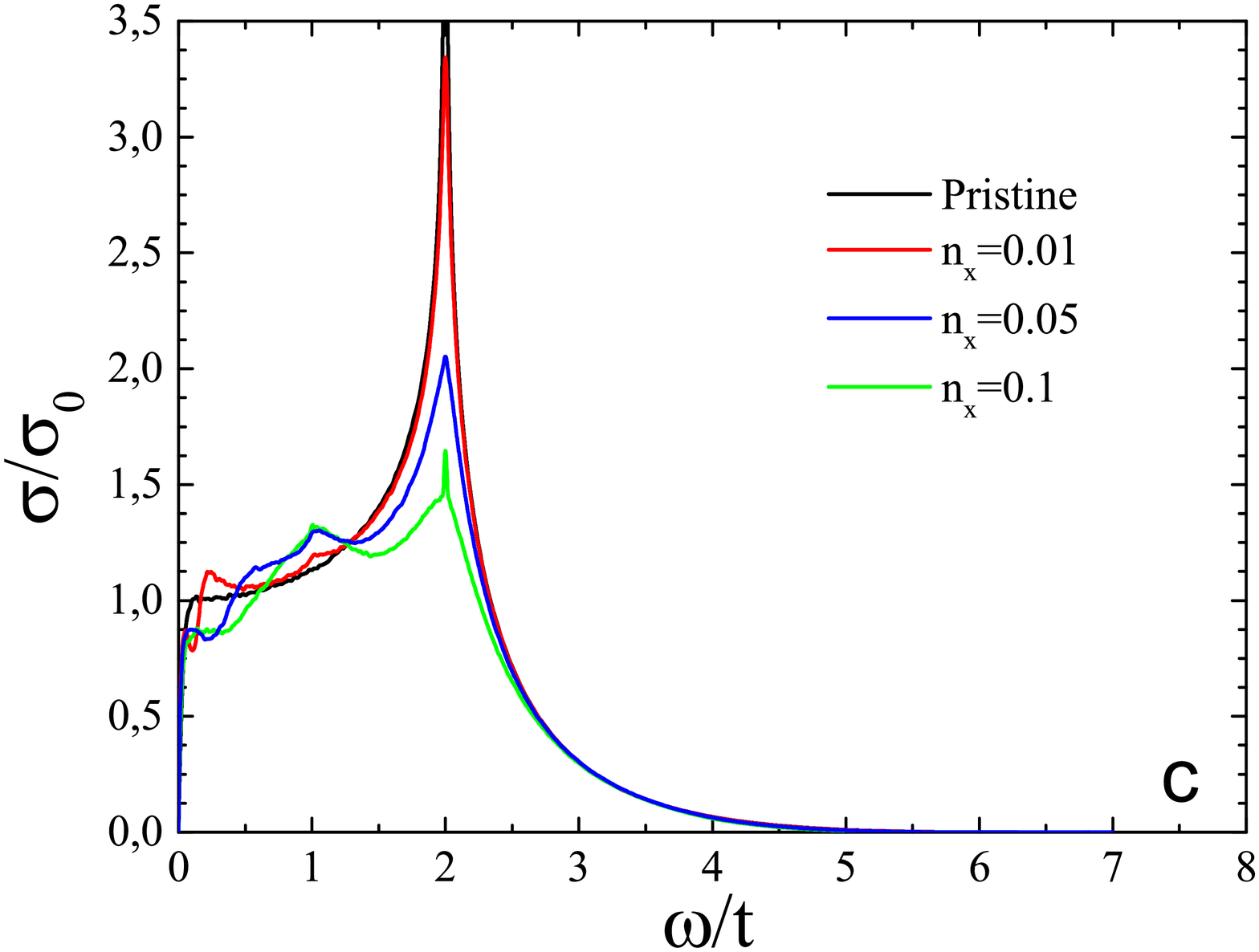}
\includegraphics[width=9cm]{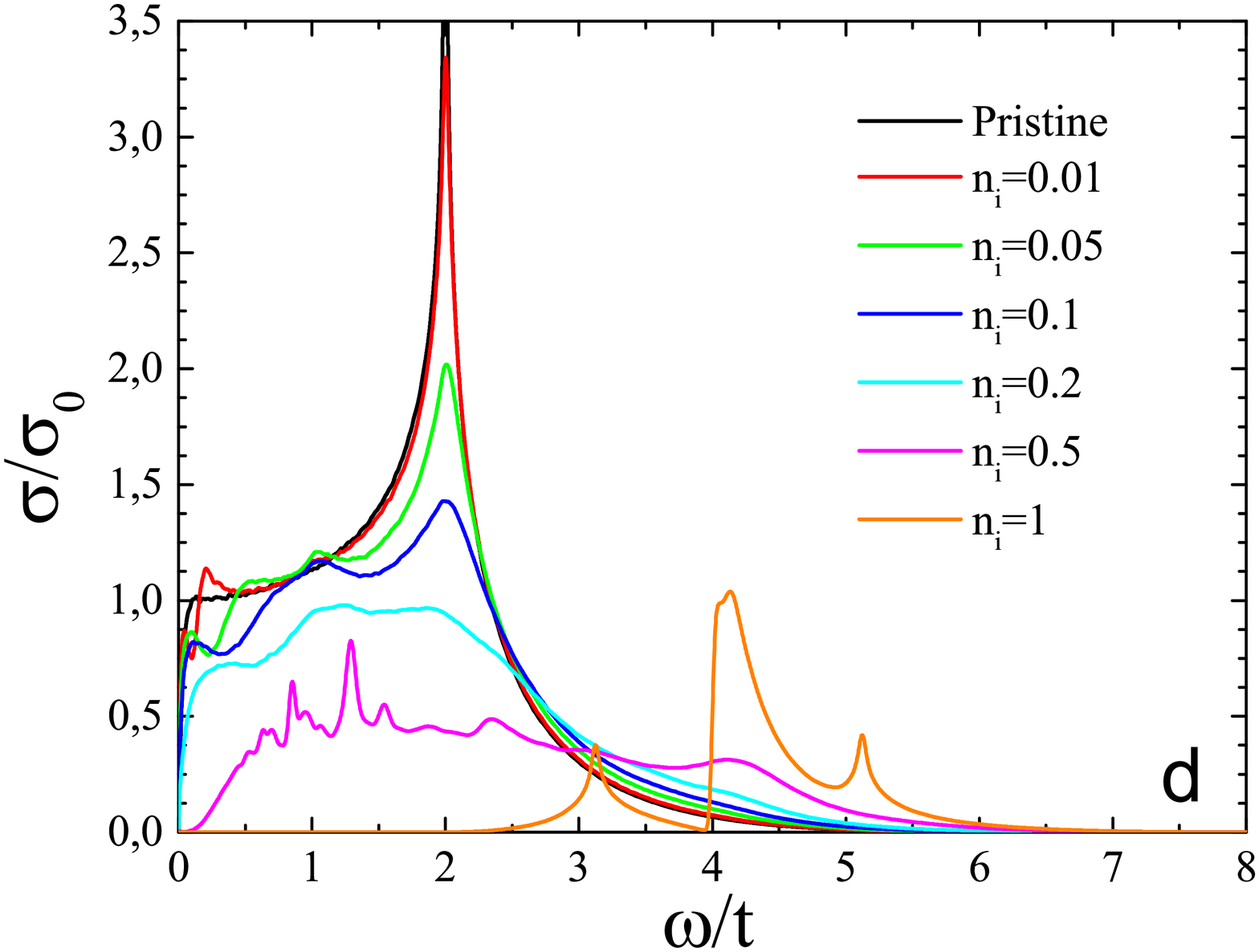}
}
\caption[]{ Optical conductivity $\sigma(\omega)$ of SLG with different kinds of disorder, in units of $\sigma_0=\pi e^2/2h$. (a) Effects of random on-site potentials on $\sigma(\omega)$. The on-site potential is let to randomly vary in the range $[-v_r,v_r]$, for $v_{r}=0$, 0.2, 0.3, 0.4, 0.5, 0.7, 1, 2.5 (in units of $t$). (b) Effects of random hopping constants on $\sigma(\omega)$. The hopping integral is let to vary randomly between $[t-t_r,t+t_r]$, for $t_{r}=0$, 0.3, 0.35, 0.4, 0.45, 0.5 (in units of $t$). (c) Effect of random vacancies (missing atoms) in the lattice, with the concentration per carbon atom $n_{x}$ taking the values 0, 0.01, 0.05, 0.1. (d) Effects of random hydrogen-like impurities (adatoms or admolecules) with the concentration per carbon atom $n_{i}$ being 0, 0.01, 0.05, 0.1, 0.2, 0.5, 1.0. For all the plots, the value of the corresponding disorder parameter $v_r$, $t_r$, $n_x$ and $n_i$ increases in order of decreasing sharpness of the peak at $\omega=2t$. 
}
\label{Fig:OpCond}
\end{figure*}

\begin{figure*}
\mbox{
\includegraphics[width=9cm]{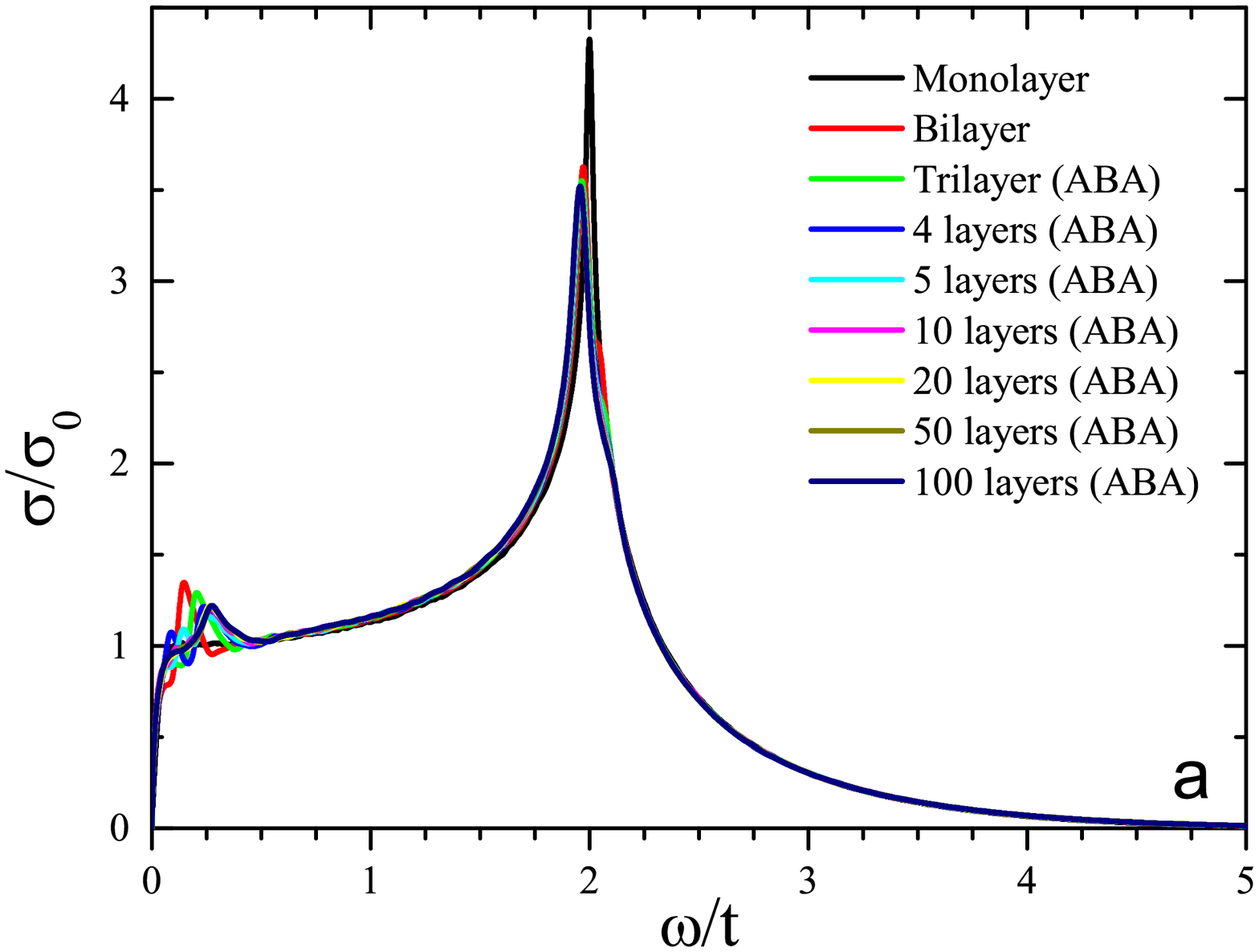}
\includegraphics[width=9cm]{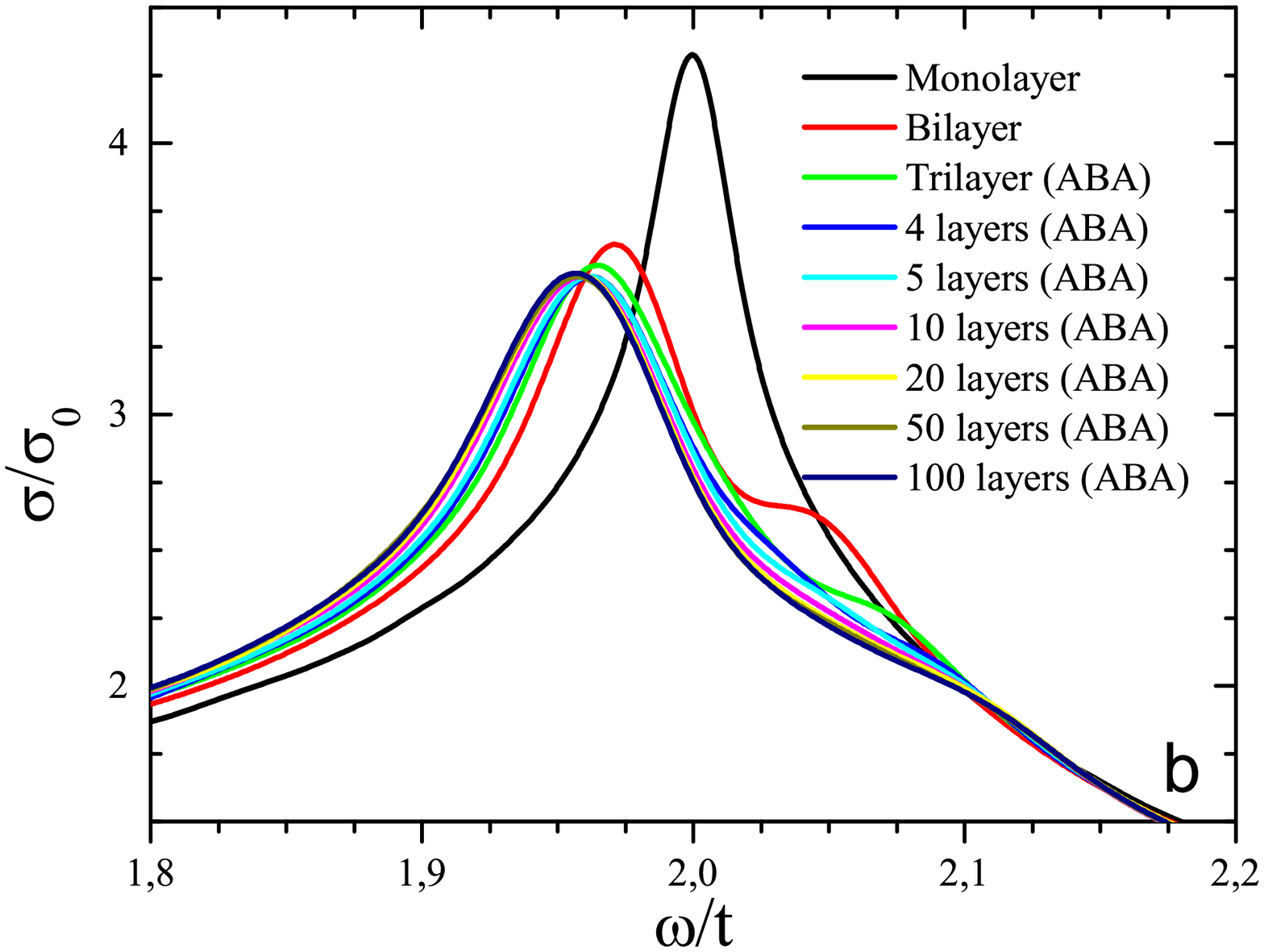}
}
\mbox{
\includegraphics[width=9cm]{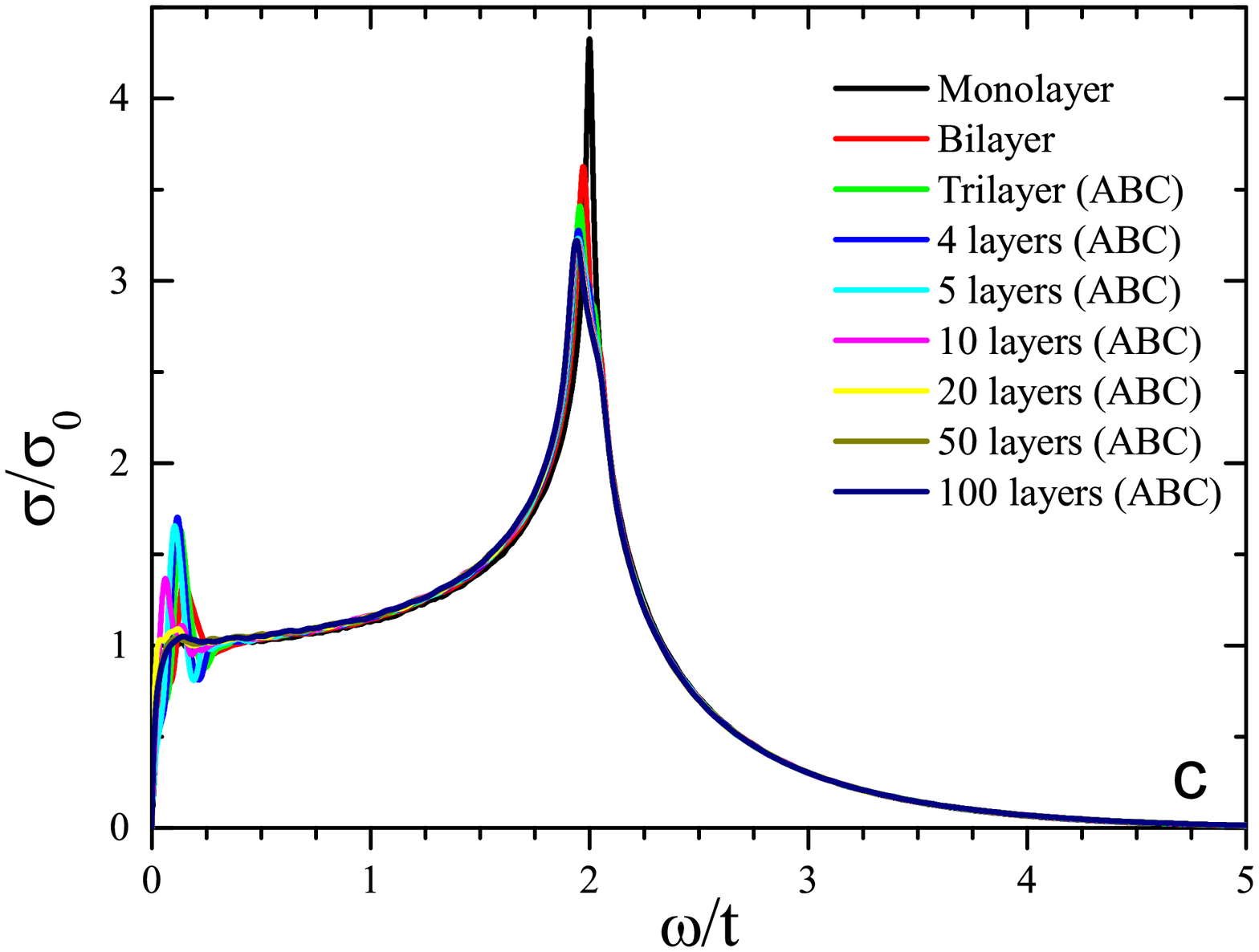}
\includegraphics[width=9cm]{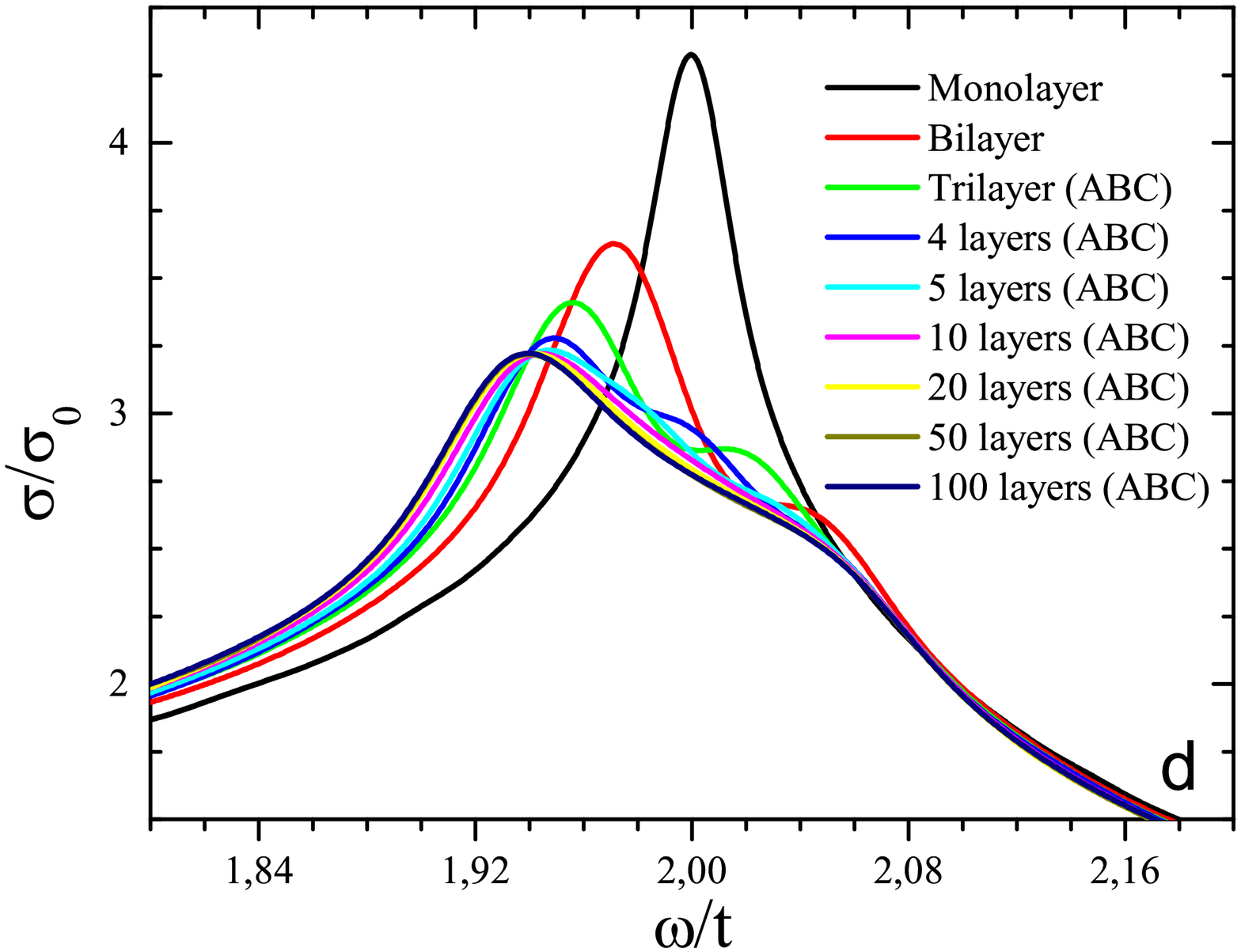}
}
\caption[]{Effect on $\sigma(\omega)$ of the number of layers and stacking order of clean multi-layer graphene. Plots (a) and (b) are for the ABA stacking, whereas plots (c) and (d) are for ABC. Plot (b) and (d) are zooms of the spectra (a) and (c), respectively, around the resonance peak at $\omega\approx 2t$. The main effect, for present purposes, is a local deformation of the peaks [zooms (b) and (d)]. The blue wing, on which the Fr\"ohlich  resonance stands, is hardly affected. The maximum separation between sub-peaks of the resonance top is determined by the coupling energy between adjacent layers, $\gamma _{1}=2t/15$.
}
\label{Fig:MLG}
\end{figure*}

The density of states is calculated by the Fourier transform of the time-dependent correlation functions (Hams and De Raedt 2000, Yuan et al. 2010)
\begin{equation}
\rho \left( \varepsilon \right) =\frac{1}{2\pi }\int_{-\infty }^{\infty
}e^{i\varepsilon t}\left\langle \varphi \right\vert e^{-iHt}\left\vert
\varphi \right\rangle dt,  \label{Eq:DOS}
\end{equation}
with the same initial state $\left\vert \varphi \right\rangle $ defined in Eq.~(\ref{Eq:phi0}). For a more detailed description and discussion of our numerical method we refer to Ref. (Yuan et al. 2010). In this paper, we fix
the temperature to $T=300$K. In our numerical calculations, we use periodic boundary
conditions in the plane ($XY$) of graphene layers, and open boundary
conditions in the stacking direction ($Z$). The size of SLG in our calculations is $8192\times 8192$ or $4096\times 4096$. For MLG, we keep the total number of atoms in the same order as for SLG.

Each kind of disorder considered here is controlled by a single parameter:
1) random on-site potential is controlled by $v_{r}$: the potential $v_{i}$ on each site can randomly change within the range $[-v_{r},v_{r}]$; 2) random hopping $t_{r}$: the hopping between two nearest neighbors can randomly vary within the range $[t-t_{r},t+t_{r}]$; 3) concentration of random distributed vacancies $n_{x}=N_x/N$, which is the probability for an atom to be replaced by a vacancy in the lattice, where $N_x$ is the total number of vacancies and $N$ is the total number of sites in the lattice; 4) similarly, the concentration of random distributed hydrogen adatoms, is controlled by the parameter $n_{i}=N_i/N$,  which is the probability for an atom to absorb a hydrogen atom.

It should be stressed that our calculations automatically account for the variation of the matrix element of the dipole moment, which is often disregarded for simplicity (see Bassani F. and Pastori-Parravicini G. 1967; Mennella et al. 1998), but is essential for our present purposes.

\section{Computational results: conductance}

The most direct test of the theory is the measurement of the absorbance of light perpendicularly through the graphene layer, which is proportional to the conductance of the latter. For this aim, in this section we present numerical results for the optical conductivity, obtained from Eq. (\ref{Eq:OpCond}), for SLG and MLG with different kinds of disorder. Our results are shown in Fig. \ref{Fig:OpCond}.

We first consider the effect of random local change of on-site potentials and random renormalization of the hopping, which correspond to the diagonal and off-diagonal disorders in the single-layer Hamiltonian Eq. \ref{Eq:H_SLG}, respectively. The former acts as a local shift of the chemical potential of the Dirac fermions, and the latter is associated to changes of distance or angles between the $p_z$ orbitals. As we have explained above, we introduce non-correlated disorders in the on-site potentials by considering that the on-site potential $v_i$ is random and uniformly distributed (independently of each site $i$) between the values $-v_r$ and $+v_r$. Similarly, non-correlated disorder in the nearest-neighbor hopping is introduced by letting $t_{ij}$ be random and uniformly distributed (independently of the pair of neighboring sites $\langle i,j\rangle$) between $t-t_r$ and $t+t_r$. Both kinds of disorder have a rather similar effect on the optical conductivity, as it can be seen from Figs. \ref{Fig:OpCond}(a) and (b). The spectrum is smeared at the resonance energy $\omega = 2t$, and the smeared region expands around the singularitiesÕ neighboring areas as the strength of the disorder increases. We remind here that the origin of this resonance is associated to particle-hole excitations between states of the valence band with energy $E \approx-t$ and states of the conduction band with energy $E \approx t$, which contribute to $\sigma(\omega)$ with a strong spectral weight due to the enhanced density of states at the Van Hove singularities of the $\pi$ bands, in agreement with recent experimental results (Mak et al. 2008).

We next consider the influence of vacancies and hydrogen impurities on the spectrum. Apart from the creation of mid-gap states in the DOS, which is not relevant for the present discussion, the presence of vacancies and/or adatoms leads also to a smearing of the Van Hove singularities in the DOS. This is in fact the behavior found in Fig. \ref{Fig:OpCond}(c) and (d), where we show the optical conductivity of SLG with different concentrations of impurities. In particular, the presence of low or moderate concentrations of hydrogen impurities, which are introduced by the formation of a chemical bond between a carbon atom from the graphene sheet and a carbon, oxygen, or hydrogen atom from an adsorbed organic molecule (CH$_{3}$, C$_{2}$H$_{5}$, CH$_{2}$OH, as well as H and OH groups), have a quite similar effect on the electronic structure and transport properties of graphene as compared to the effect of vacancies. In our work, the adsorbates are described by the Hamiltonian $H_{imp}$, Eq. (\ref{Eq:Himp}), where the band parameters $V \approx 2t$ and $\varepsilon_d \approx -t/16$ are obtained from ab-initio density functional theory (DFT) calculations (Wehling et al. 2010). On the other hand, large concentrations of hydrogen impurities change dramatically the optical properties of the system. In fact, for the maximum concentration $n_i=1$, which correspond to an H atom bonded to each C atom of the lattice, we have no longer graphene but a completely different material, called {\it graphane} which contrary to graphene, has a band gap of $\Delta \approx 2t$. Therefore, the optical conductivity at low frequencies is zero from $\omega=0$ to $\omega=2t$, and from this energy it grows smoothly, presenting a series of peaks associated to newly created Van Hove singularities in the spectrum, as it can be seen by looking at the purple line of Fig. \ref{Fig:OpCond}(d). However, the more realistic case of low or moderate concentrations of hydrogen, leads to a smearing of the Van Hove singularity with a transfer of spectral weight to new peaks in the spectrum, particularly to a new peak at $\omega\approx t$ which is associated to transitions between the impurity zero energy band and the Van Hove singularities at $E=t$.

As an infinite sheet of graphene proper cannot survive in vacuum (it folds over itself), we have explored the electrical effects of stacking an increasing number of sheets. In equilibrium, two stacking schemes must be considered, according to the arrangement of the benzenic rings facing each other: ABA and ABC (see Fig. \ref{Fig:Stacking}). The inter-layer distance, experimentally measured, has a value of $d\approx 3.36$~\AA. The inter-layer hoppings considered here are $\gamma _{1}=2t/15$ and $\gamma _{3}=t/10$, the same values as in graphite, and are also sketched in Fig. \ref{Fig:Stacking}. The stacking order, which leads to different low energy properties of ABA and ABC MLGs, has very limited effect on the high energy spectrum and, in particular, on the blue wing of the $\pi$ resonance, where the
 Fr\"ohlich bump stands. This can be seen in Fig. \ref{Fig:MLG}(a) and (c), which show the full $\sigma(\omega)$ spectrum of MLG with an increasing number of layers, up to 100 layers, with ABA- and ABC-stacking order, respectively. The UV bump, therefore, will not be affected.  By contrast, the effect on the peak is remarkable, and is a classical outcome of coupling identical oscillators: creating an increasing number of subpeaks on both sides of the single-oscillator peak. Note, however, that their excursion from the latter does not exceed an amount determined by the coupling energy assumed between layers (see Gr\"uneis et al. 2008), $\sim3\%$ in the present case. As a consequence, the peak becomes less sharp. This must be a major reason why the $\pi$ and $\sigma$ resonances of graphite are much less peaked than those of graphene (see also Kobayashi and Uemura 1968; Johnson and Dresselhaus 1972).

Summarizing, clean graphene has the steepest slope of the $\pi$-feature. This slope is shown below to govern the width of the Fr\"ohlich  resonance; so, we will concentrate on those types of disorder which have the strongest effect on the slope of the blue wing. The most striking aspect of these curves, for our present purposes, is the continuous variation of the slope of the high energy wing of the $\pi$ resonance, as the disorder parameter increases. All types of disorder exhibit this behavior, albeit to different extents. Other, so-called correlated disorders were also studied, but they do not affect the blue wing notably. The reason is that for correlated disorder, for which the distribution of the disorder follows particular topological structures, such as Gaussian potentials or Gaussian hopping parameters, or resonant clusters with groups of vacancies or hydrogen adatoms, the electronic DOS of clean graphene is highly preserved. Therefore, a SLG or a MLG sample with a given amount of impurities, will present an optical spectrum closer to that of clean graphene (in particular presenting a prominent peak at the $\pi$-resonance) if the impurities are correlated, rather than if they are spread in a non-correlated manner in the sample. Another remarkable effect of disorder is the progressive damping of the $\pi$ resonance itself, which is a very well experimentally documented phenomenon exhibited by families of a-C:H materials. We emphasize that no {\it ad hoc} broadening mechanism, such as an arbitrary life time, is invoked in the present treatment. Each single parameter controlling any given type of disorder can be estimated from the structural and environmental characteristics of the corresponding sample.

\section{The dielectric function}

The study of the Fr\"ohlich  resonance requires the knowledge of the dielectric function. The imaginary part of the relative dielectric function, $\epsilon= \epsilon_{1}$+i$\epsilon_{2}$ is related to the conductance computed above by
\begin{equation}
\epsilon_{2}=\frac{\sigma}{\omega},
\end{equation}
where $\sigma$ and $\omega$ are in units of s$^{-1}$ in the SI system. While the conductance $\sigma$ is completely defined in terms of $\sigma_{0}=\pi e^{2}/2h$, which depends only on fundamental constants, the corresponding energy variable is
 given only to a factor $t$, which represents the coupling energy between neighboring atoms in perfect, clean graphene. The latter is not known with great accuracy, and the value adopted for it differs according to authors (see, for instance, Gr\"uneis et al. 2008, Kobayashi J. and Uemura Y. 1968). Here, we take $2t=4.21$ eV, so that the $\pi$ peak of the dielectric function falls near the measured position for graphite (Taft and Philipp 1965; Greenaway et al. 1969; Tosatti and Bassani 1970).

The higher energy $\sigma$ resonance of graphite peaks near 14 eV but its red wing slightly overlaps the blue wing of the $\pi$ resonance, in between them, and this bears upon the position of the Fr\"ohlich  resonance. However, the theoretical procedure outlined in Sec. 2 has not yet been extended to the $\sigma$ resonance (see Sec. 7.7). We were therefore led to use, instead, the values of $\epsilon_{2}$ deduced by Taft and Philipp \cite{tp} from reflectance measurements between 9 and 32 eV, and joined them continuously to each of the $\pi$ curves we computed as explained above. By extending the overall frequency range, this also helps improving the accuracy of the Kramers-Kronig transformation which was applied to $\epsilon_{2}$ in order to deduce $\epsilon_{1}$.

 This $anzatz$ is justified by the fact that the intensity of the $\sigma$ resonance of carbonaceous materials is largely decoupled from that of the $\pi$ resonance, as the two singularities are separated by more than 6 eV; a fact which is likely to be related with the lack of correlation, in the interstellar extinction curves, between the UV bump and the far UV curvature.  Moreover, disorder is not likely to perturb excessively the $\sigma$ resonance since, as recalled in the Introduction, the latter is associated with the $pp\sigma$ type of bonding which survives the transition to $sp^{3}$ hybridization that dominates structurally disordered materials. This statement is borne out by the comparison of the VUV spectra of graphite with several species of coals of different evolutionary states, which correspond to different degrees of disorder Papoular et al. 1995).

Figures 4 to 6 exhibit some of the results for 3 types of disorder, excluding the vacancy disorder, which was shown above not to affect
significantly the slope of the blue wing of the conductance resonance. They mirror the curves of Fig. \ref{Fig:OpCond} (a), (b) and (d). The general trends of interest to us in the three graphs are the same, even though their magnitudes may differ: the peak for pure and perfect graphene is a logarithmic singularity; as disorder increases, it is progressively damped, and ultimately drowns into a continuum; concurrently, the blue slope becomes less and less steep. The former behavior may be ascribed to the progressive suppression of $sp^{2}$ bonds, and accounts for the observed difference between graphite and various species of a-C:H (see, Robertson 1986, Mennella et al. 1998) or coals (see Papoular et al. 1995). The change in slope is shown below to account for the width variations of the Fr\"ohlich  resonance feature. Note that this effect is all but absent in the case of vacancy disorder.

\begin{figure}
\resizebox{\hsize}{!}{\includegraphics{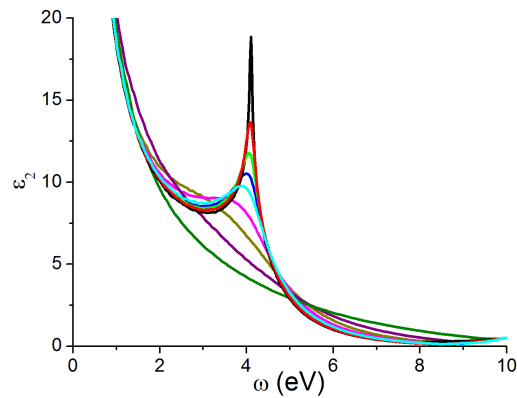}}
\caption[]{Effects of random on-site potentials on the dielectric function. The disorder parameter, $v_{r}$ (in units of $t$), is 0, 0.2, 0.3, 0.4, 0.5, 0.7, 1, 1.5 2.5, in order of decreasing sharpness of the resonance peak.}
\label{Fig:eppvrB}
\end{figure}

\begin{figure}
\resizebox{\hsize}{!}{\includegraphics{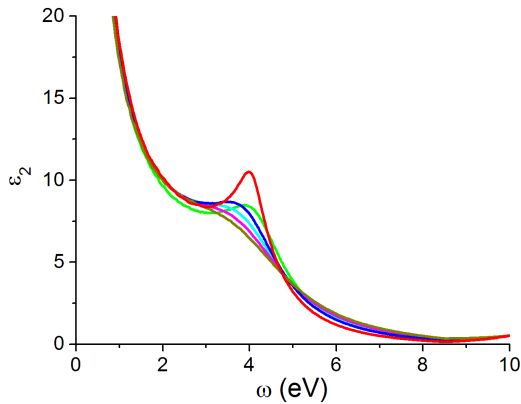}}
\caption[]{Effects of random hopping constants on the dielectric function. The disorder parameter, $t_{r}$ (in units of $t$), is 0.2, 0.3, 0.35, 0.4, 0.45, 0.5, in order of decreasing sharpness of the resonance peak. }
\label{Fig:epptrB}
\end{figure}

\begin{figure}
\resizebox{\hsize}{!}{\includegraphics{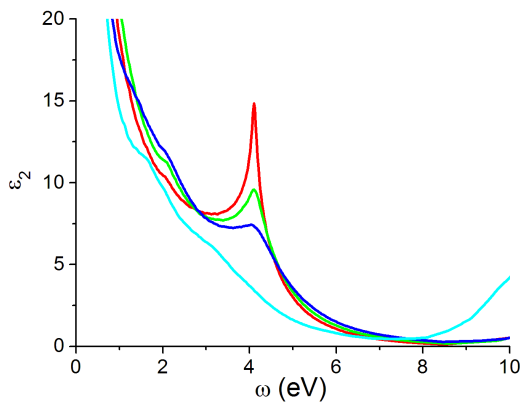}}
\caption[]{Effects of random atomic impurities on the dielectric function. The disorder parameter, $n_{i}$, is 0.01, 0.05, 0.1, 0.2, in order of decreasing sharpness of the resonance peak. }
\label{Fig:eppniB}
\end{figure}

\section{Application to the 2175-\AA~ feature}

\begin{figure}
\resizebox{\hsize}{!}{\includegraphics{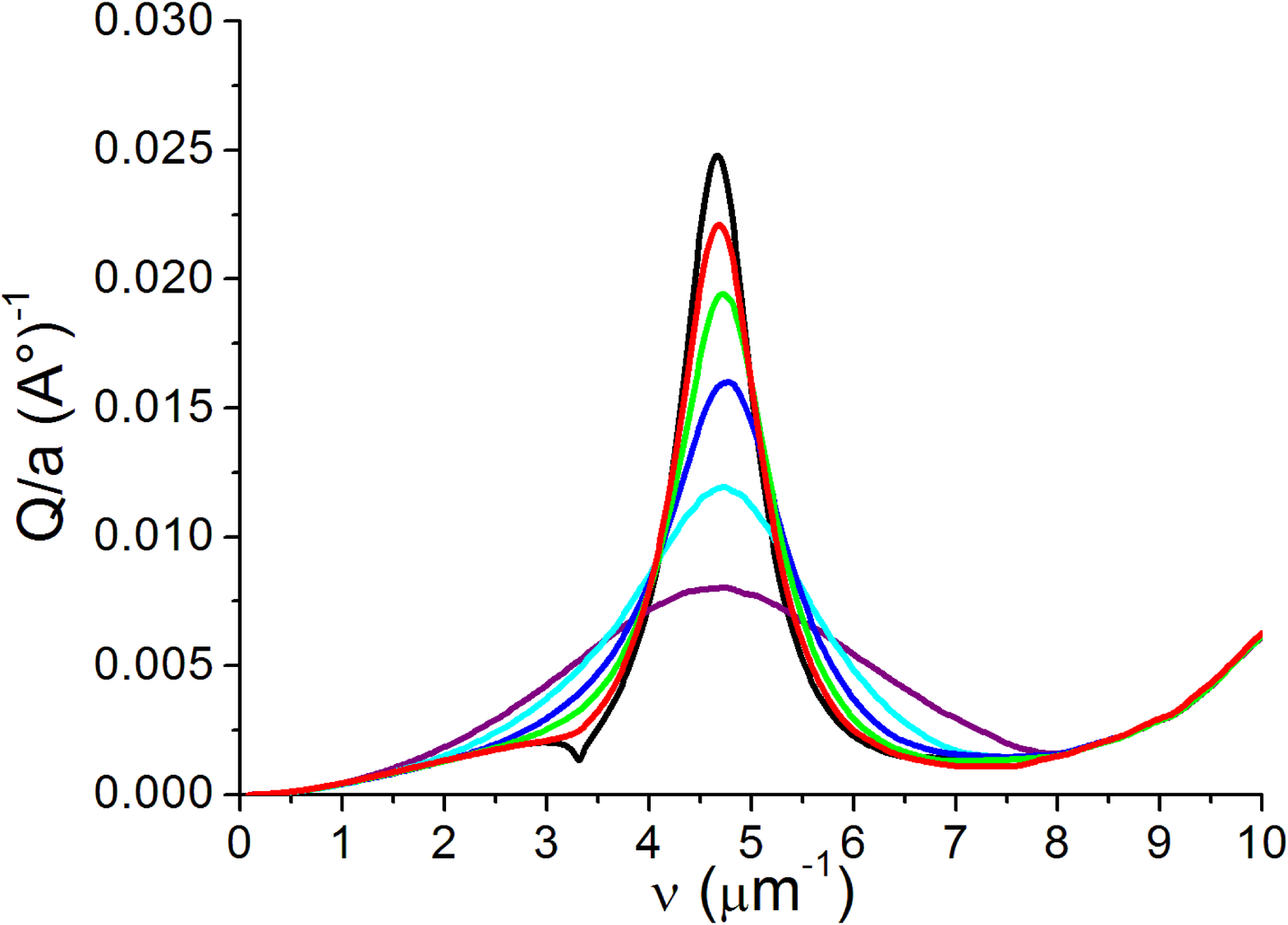}}
\caption[]{Effects of random on-site potentials on the Fr\"ohlich  resonance of the $\pi$ resonance. The disorder parameter, $v_{r}$ (in units of $t$), is 0, 0.4, 0.7, 1, 1.5, 2.5, in order of decreasing sharpness of the resonance peak.}
\label{Fig:QtvrB}
\end{figure}

The weak scattering apparently included in the astronomical UV bump extinction, elicits the assumption of small particles for the band carrier, and, hence, the application of the Rayleigh approximation. Consequently, surface effects cannot be neglected, which entails the use of the Fr\"ohlich expression for the extinction efficiency of a small sphere of radius $a$:
\begin{equation}
Q/a=\frac{24 \pi}{\lambda}\frac{\epsilon_{2}}{(\epsilon_{1}+L^{-1}-1)^{2}+(\epsilon_{2})^{2}}
\end{equation}
where $\lambda$ is the wavelength of the radiation, $\epsilon_{1,2}$ are, respectively, the real and imaginary parts of the dielectric function of the assumed homogeneous material; $L$ is the shape parameter of the supposedly ellipsoidal grain, 1/3 in the case of a sphere (see Bohren and Huffman 1983); $a$ is the radius of the sphere. In this approximation, there is no need to distinguish between absorption and extinction efficiencies, both being designated by $Q$. The real part, $\epsilon_{1}$, of the dielectric function is obtained by applying the Kramers-Kronig transformation to $\epsilon_{2}$ from 0 to 32 eV. For this operation, Papoular \cite{rjp} built a transform algorithm based on the Maximum Entropy method. Figures 7 to 9 show the results of computations for a few representative cases. The ``clean'' case (all disorder parameters null) is represented in Fig. \ref{Fig:QtvrB} only (black line); its peak is slightly higher than that for $n_{i}=0.01$ in Fig. \ref{Fig:QtniB}.

 As expected, as disorder increases, the width of the feature increases and the height decreases, both monotonously; but the peak position hardly changes from $\sim4.6\,\mu$m$^{-1}$(because of the Kramers-Kronig relation between $\epsilon_{1}$ and $\epsilon_{2}$). The minimum width and maximum height are carried by the clean case. For this case, the peak value of $Q/a$ is $\sim0.02\,\AA^{-1}$ (upper limit of $Q/a$ in our study) and the width, 
$\sim0.85\,\mu$m$^{-1}$ (lower limit in the present work). These limits are set by the logarithmic slope of the resonance (see van Hove 1953) .

Strictly speaking, these results apply to a single graphene layer. However, they derive solely from the behavior of the blue wing of the conductance resonance, and this was shown to be insensitive to the number of stacked layers [Fig. \ref{Fig:MLG} (a) and (c)], whatever the stacking scheme. Figure 10 illustrates this point. We therefore assume, below, that they apply equally to MLG (multi-layer graphene).

\begin{figure}
\resizebox{\hsize}{!}{\includegraphics{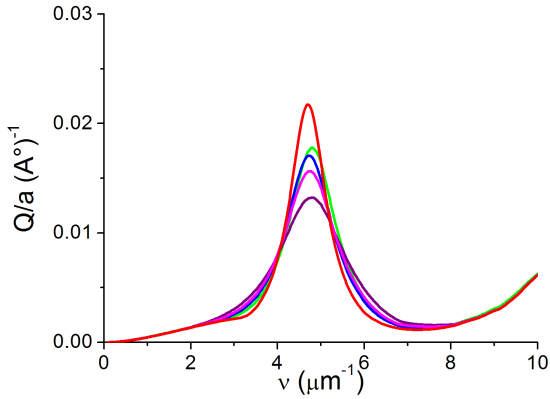}}
\caption[]{Effects of random hopping constants on the Fr\"ohlich resonance of the $\pi$ resonance. The disorder parameter, $t_{r}$ (in units of $t$), is 0.2, 0.3, 0.35, 0.4, 0.45, 0.5, in order of decreasing sharpness of the resonance peak. }
\label{Fig:QttrB}
\end{figure}

\begin{figure}
\resizebox{\hsize}{!}{\includegraphics{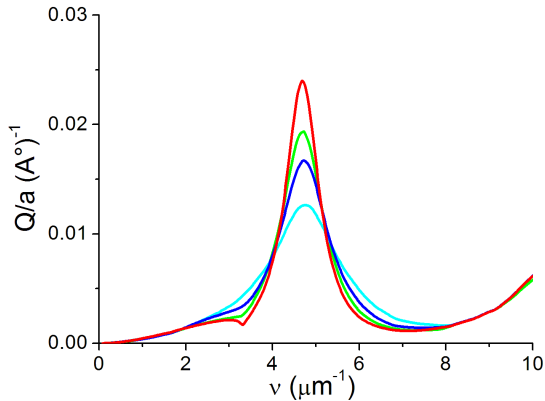}}
\caption[]{Effects of random atomic impurities on the Fr\"ohlich resonance of the $\pi$ resonance. The disorder parameter, $n_{i}$, is 0.01, 0.05, 0.1, 0.2, in order of decreasing sharpness of the resonance peak. }
\label{Fig:QtniB}
\end{figure}

\begin{figure}
\resizebox{\hsize}{!}{\includegraphics{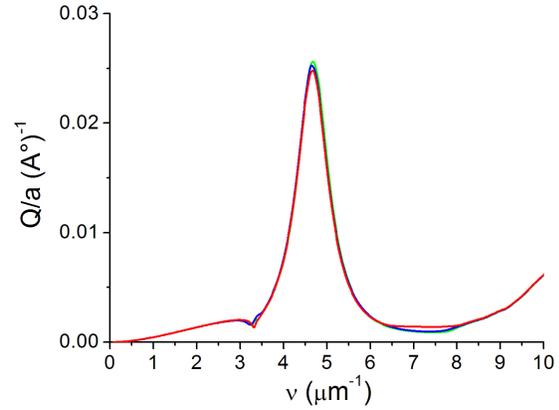}}
\caption[]{The Fr\"ohlich resonance of the $\pi$ resonance in the clean case for a single layer (red), a bi-layer (green) and an ABA stack of 100 layers (blue).}
 \label{Fig:fig_Q_SLG_MLG}
\end{figure}

\section{Application to Astronomy}

\begin{figure}

\begin{center}
{
\begin{minipage}{.2\textwidth}
  \includegraphics[width=.6\textwidth]{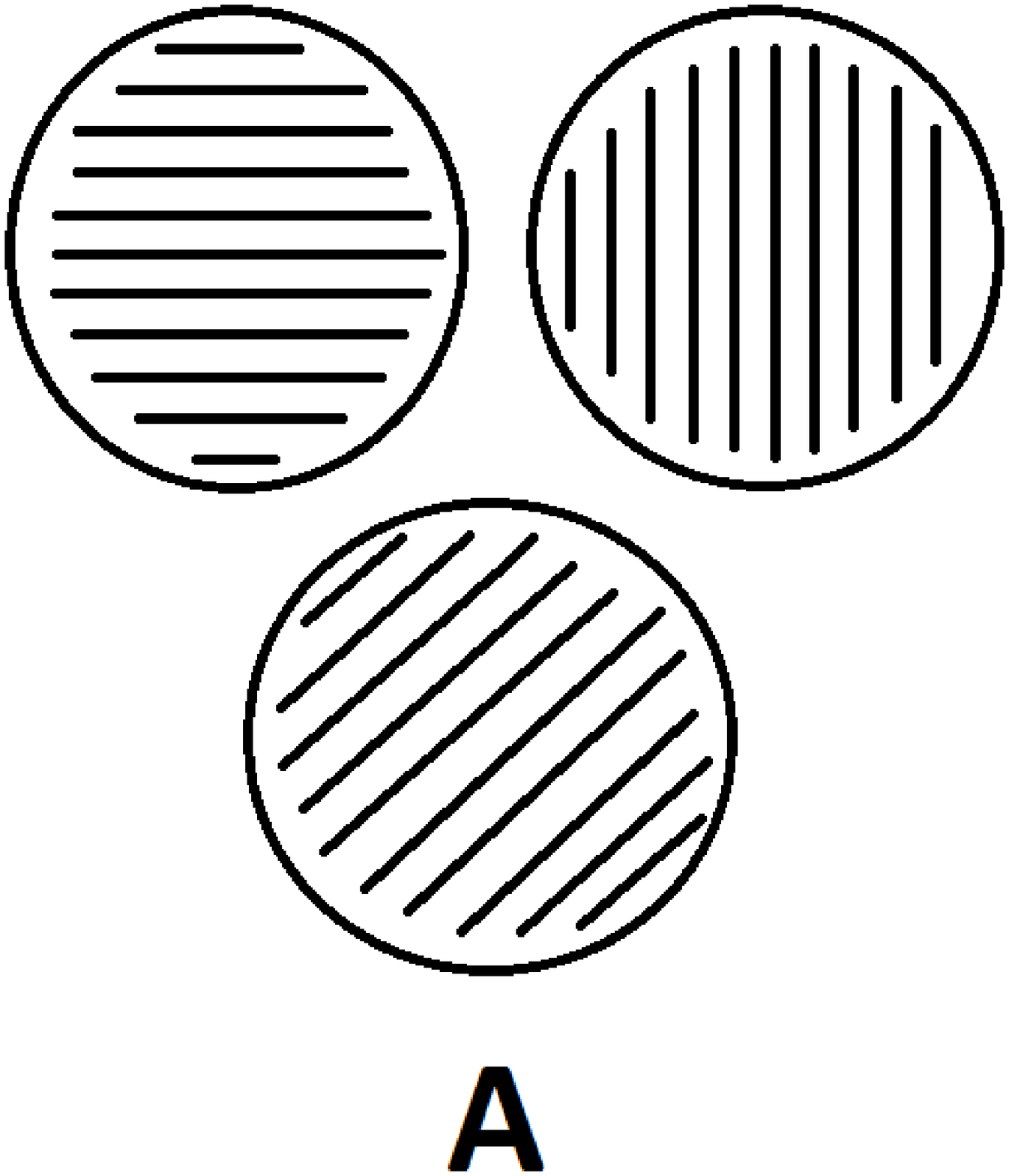}
 \label{Fig:Figure_dustA} 
  
\end{minipage}
\begin{minipage}{.2\textwidth}
 \includegraphics[width=.6\textwidth]{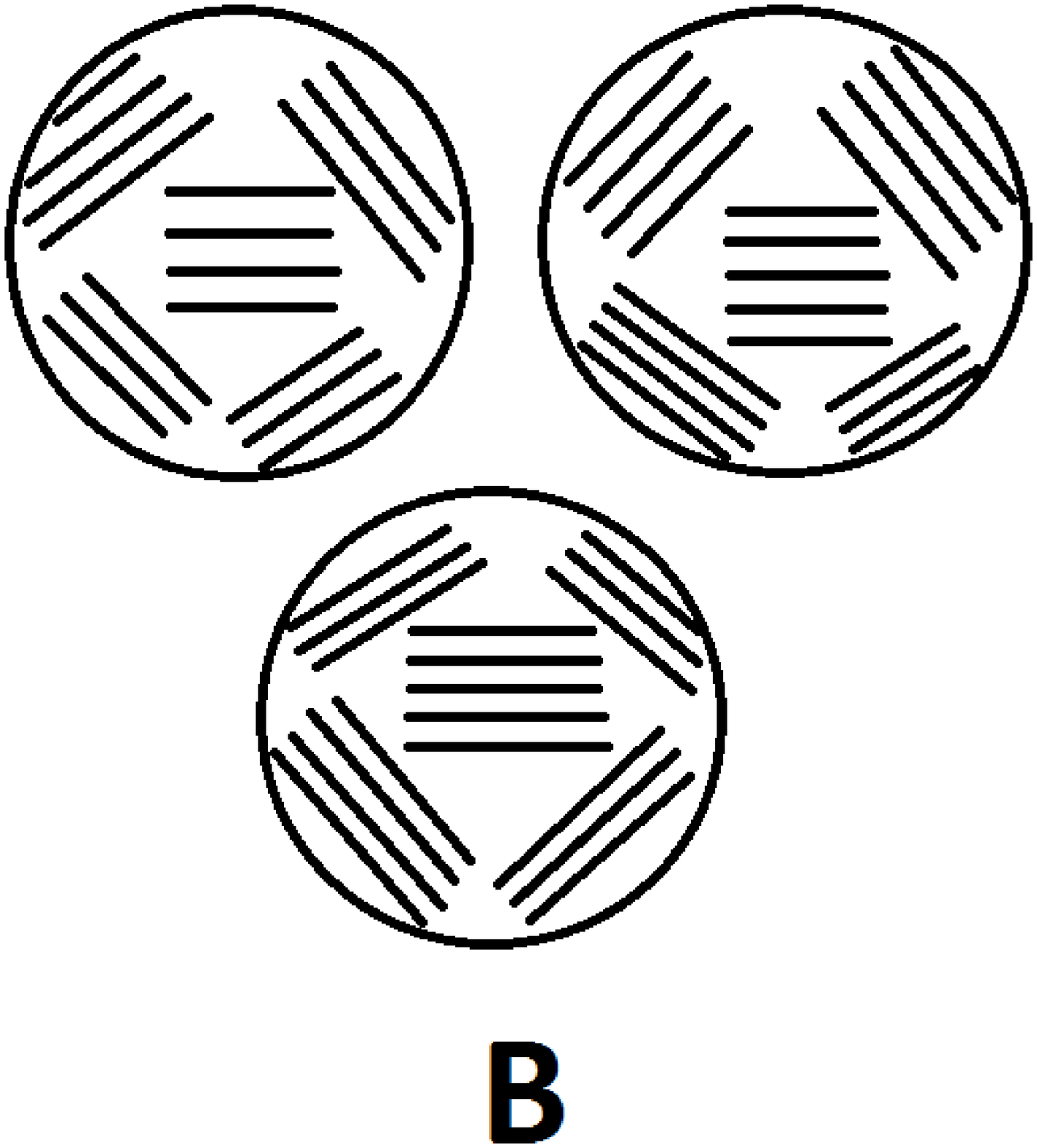}
 \label{Fig:Figure_dustB}
\end{minipage}
}
\end{center}
\caption{Models of dust particles. A: free-flying bricks ; each dust particle is made of one ordered stack of graphene layers. B: agglomerated bricks; each dust particle includes several such ``bricks''.}
\end{figure}

\begin{figure}
\resizebox{\hsize}{!}{\includegraphics{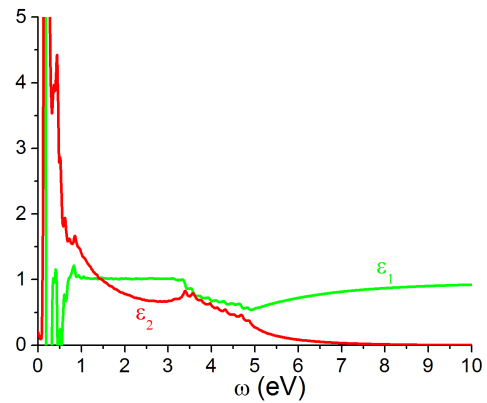}}
\caption[]{Dielectric function for the clean case, and for $\vec E\|\vec c$. Red line: imaginary part $\epsilon_{2}$; green line: real part $\epsilon_{1}$. Color on line.}
\label{Fig:epspar}
\end{figure}

Consider first the free-flying brick model (Fig. \ref{Fig:Figure_dustA}), where each dust particle is a spheroidal assembly of nano-sized portions of graphene, stacked according to the ABA or ABC scheme. Their orientation being random, the apparent extinction along a given line of sight will be the average of extinction for $\vec E\perp\vec c$ (electric field parallel to the layer planes) and for $\vec E\|\vec c$ (electric field perpendicular to these planes), with relative weights 2/3 and 1/3, respectively. We have, therefore, applied the same theoretical procedure to compute the dielectric function in the parallel case.  As an example, Fig. \ref{Fig:epspar} plots the real and imaginary parts of this function for the clean case.

\begin{figure}
\resizebox{\hsize}{!}{\includegraphics{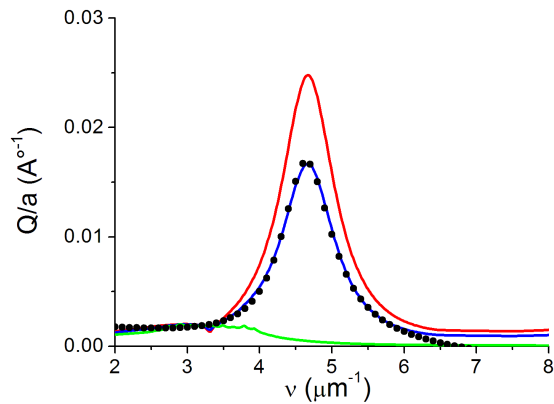}}
\caption[]{Blue line: average extinction bump for randomly oriented spheroidal stacks of clean graphene layers. Black points: Drude fit. Red upper line: extinction bump for $\vec E$ in plane. Green lower line: extinction bump for $\vec E$ normal to the plane. Color on line.}
\label{Fig:Qmoy}
\end{figure}

Correspondingly, Fig. \ref{Fig:Qmoy} plots the parallel and perpendicular extinction efficiencies for the clean case, together with their weighted average. Obviously the position and relative weakness of the ``parallel bump'' are such that the average curve differs from the ``perpendicular bump'' only in that its height is reduced by a factor 2/3. For purposes of quantitative comparison, the average $Q/a$ was fitted with  a ``Drude'' curve, as initiated by Fitzpatrick and Massa \cite{fm86}, also shown in the figure. It peaks at 4.65 $\mu$m$^{-1}$ and its width is $\gamma\sim0.85\,\mu$m$^{-1}$. Only for highly disordered structures, when the ``perpendicular'' bump is weakened beyond the range considered in Fig. 7-9, does the effect of the ``parallel'' bump on the average become notable.

 \begin{table}
 \caption[]{Experimental model results.}
\begin{flushleft}
\begin{tabular}{lll}
\hline
Name & $\nu_{0}$ & $\gamma$\\
\hline
AR$^{a}$ & 4.46 & 1.48\\
\hline
BCC V $^{b}$ & 4.7 & 1.6\\
\hline
Free bricks$^{c}$ & 4.56 & 1.15\\
\hline
PCG agglom$^{d}$ & 4.52 & 1.3\\
\hline
\end{tabular}
\end{flushleft}
\begin{list}{}{}
\item The units of $\nu_{0}$ and $\gamma$ above are $\mu$m$^{-1}$.
\item [a]Schnaiter et al. \cite{sch}
\item [b]Mennella et al. \cite{men98}
\item [c,d]Papoular and Papoular \cite{pap09}
 \end{list}
\end{table}

In another dust model (Fig. 11B), the bricks are agglomerated into a single particle: industrial polycrystalline graphite (PCG) is a case in point (Papoular and Papoular 2009). It is made of powder of nano-sized highly graphitized and pure carbon grains, compressed under high pressure. The measured reflectance of PCG was modeled by 2 pairs of Lorentzian oscillators, for the parallel and perpendicular $\pi$ and $\sigma$ dielectric functions. The isotropic dielectric function is a mix of the parallel and perpendicular dielectric functions, and was computed using the Bruggeman mixing formula. The extinction feature was then found to peak at 4.52 $\mu$m$^{-1}$, with FWHM 1.3 $\mu$m$^{-1}$ (``PCG agglom'' in Tab. 1). Table 1 collects examples of relevant experimental results, also symbolized in Fig. 14.

It is instructive to envision the behavior of this material if each of its constitutive graphitic bricks were set free in the atmosphere, as in model 10 A. The parallel and perpendicular extinctions in that case were calculated using the same dielectric functions as deduced above from the reflectance of PCG. Their weighted average (1/3, 2/3) was found to exhibit a bump at 4.56 $\mu$m$^{-1}$, with FWHM 1.15 $\mu$m$^{-1}$. Obviously, agglomeration of the bricks entails some broadening as do the various types of disorder considered above. There is also some redshift. Unlike with the free-flying bricks, the effective dielectric function, here, is very sensitive to the parallel dielectric functions.

 A similar behavior is displayed by  nano-sized HAC particles agglomerated in pure argon atmosphere (Schnaiter et al. 1998): the feature width increases with the degree of clustering (their Fig. 3; the narrowest of these features is included as``AR'' in Tab. 1 and symbolized in Fig. 14). Agglomerates obtained with hydrogen included in the atmosphere show less clear trends.

\section{Discussion}

\subsection{Comparison with observations}

\begin{figure}
\resizebox{\hsize}{!}{\includegraphics{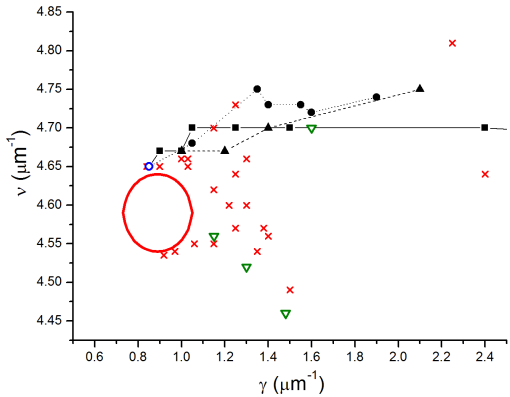}}
\caption[]{Peak position vs FWHM of the UV bump. {\it Observations} (adapted from Fig. 16 of Fitzpatrick and Massa 2007): the red ellipse encloses the vast majority of observed representative points; red crosses: a few outliers. {\it Our computations}: Open circle: randomly oriented clean and ordered graphene bricks. Filled squares and solid line: random on-site potentials ($v_{r}=0$, 0.2, 0.4, 0.5, 0.7, 1, 1.5, 2.5, 5); filled circles and dotted line: random hopping parameters ($t_{r}=0.1$, 0.2, 0.3, 0.35, 0.4, 0.45, 0.5); filled triangles and dashed line: random foreign atoms (e.g. hydrogen; $n_{i}$=1, 5, 10, 20 $\%$). {\it Laboratory models}: open triangles (see Table 1). Color on line.}
\label{Fig:FMfig16}
\end{figure}

All cases computed above are represented in Fig. \ref{Fig:FMfig16} as points with coordinates ($\gamma,\nu_{0}$). For comparison, the points plotted by Fitzpatrick and Massa \cite{fm07} in their Fig. 16 are also represented (schematically). While there is a rough agreement in the width distribution (abscissae), the ordinates, $\nu_{0}$, of the computed points are systematically shifted upwards by $\sim0.05$~\AA, or about 1 \%. $\nu_{0}$ depends slightly on the fundamental hopping parameter $t$ (Sec. 2), and mostly on the electron plasma frequency of the system, which, in turn, depends on the shape of the computed electronic bands through the effective number of electrons, $n_{eff}$ (see Taft and Philipp 1965). In the tight binding scheme adopted here, this is found to be 0.56 electron/atom, a value which is comparable to that determined experimentally for graphite, by applying the $\epsilon_{2}$-Sum Rule (see Altarelli et al. 1972). According to the latter, $n_{eff}$ is linearly related to $\epsilon_{2}$.

  Calculation shows that $\nu_{0}$ can be brought down to the right value by taking $n_{eff}=0.53$, which amounts to reducing $\epsilon_{2}$ uniformly by 6 \%. Although this is well within experimental errors of measurement of
 $\epsilon_{2}$, we chose to keep $n_{eff}=0.56$, so as to let one assess the validity of the tight binding assumption. As for the minimum width determined here, 0.85 $\mu$m$^{-1}$, we note that it is slightly higher than the value $\sim0.8 \,\mu$m$^{-1}$ suggested by astronomical measurements. This is perhaps one weakness of our model, or of the tight-binding approximation used in the derivation.

On the other hand, it should be stressed that, apart from the selected disorder type and parameter, which define the bump carrier structure, the only adjustable parameter in the algorithm is the nearest-neighbor hopping parameter, $t$ (half the transition energy at resonance). Even this is restricted by experiment within tight limits. This ensures that the relation between disorder and bump properties remains clear: disorder damps the resonance, which flattens its high-energy wing; hence, the UV bump is weakened and broadened. The reason why the peak position does not change much lies in the Kramers-Kronig relation between $\epsilon_{1}$ and $\epsilon_{2}$.

 Figure \ref{Fig:FMfig16} shows that, for moderate degrees of disorder, the representative points flock near the clean case (minimum bump width), in about the same width range as the observed widths. This sets upper limits to the extent of disorder suggested by the present model for bump carriers. Estimates of these limits for the parameters of the three  types of disorder considered here are, respectively: $v_{r}\sim0.7t$, $t_{r}\sim0.3t$ and $n_{i}\sim7\%$.

 For higher values of $v_{r}$ (not shown in Fig. \ref{Fig:FMfig16}) it is found that the peak height and wavenumber decrease, and the width increases, with disorder. This is in line with the measurements on coals in order of decreasing graphitization (see Papoular et al. 1993). Figures 4 to 6 for $\epsilon_{2}$ also suggest that highly disordered grains provide a UV continuum, as expected for strong resonance damping, also in accordance with the measurements on coals. Such grains may contribute to the ISEC continuum.

Mennella et al. \cite{men98} obtained HAC samples from electrical discharges and measured their extinction efficiency, from which they deduced the dielectric function and the Fr\"ohlich resonance. Even the $\pi$ resonance hardly emerges in nascent samples. After thermal annealing, it is distinctly visible, and the UV bump shows up. Under further treatment by UV irradiation, the latter becomes narrower and stronger; this is included in Tab. 1 as BCC V and symbolized in Fig. \ref{Fig:FMfig16}. In order to obtain still narrower features the authors extrapolated the dielectric function as a function of radiation dose, and the correspondingly computed widths were then compatible with observed bump widths, with only a slight decrease in peak wavenumber. This trend is analogous to our findings. It is attributed to an increase in aromatic cluster number and average size, and a reduction of size dispersion, which we interpret as reduced disorder.

Several other causes of bump broadening have been considered in the past. Sorrell \cite{sor}, for instance, argued that electron-phonon scattering (``viscous flow'') and collision of electrons with adsorbed H atoms limited the bump width from below. The latter phenomenon is a close kin to our impurity disorder, while the former is not considered here for we are only interested in hopping electrons, not free (conduction) ones. Another cause may be the smallness of the grain size.

The argument developed above hinges upon the graphene properties. For present purposes, this was also shown above to apply to stacks of single-layered graphene (MLG). However, graphene theory assumes an infinite plane, while, for our extinction calculations to apply, the bump carrier size is limited to $\sim100$~\AA~ by the Rayleigh condition for 2175 \AA{\ }. In fact, it can be shown that the minimum size of a graphene layer for the properties of an ideal layer to be recovered is $\sim15$~\AA, the size of a big PAH (see Sec. 7.3).

 While other causes may coexist in most astronomical instances, our model excludes their contribution in cases corresponding to the minimum observed bump widths, as the clean case already has this minimum width.

\subsection{Carbon budget}

 We seek now to estimate the fraction of the available carbon that must be locked in this model bump carrier for it to account for astronomical observations. Fitzpatrick and Massa \cite{fm07} have cast their interstellar extinction measurements in the form of a normalized color excess (in units of magnitude)
\\
$k(\lambda-V)=\frac{E(\lambda-V)}{E(B-V)}=c_{1}+c_{2}x+c_{3}D(x;\gamma,x_{0})+c_{4}F(x),$ 
\\
$D(x;\gamma,x_{0})=\frac{x^{2}}{(x^{2}-x_{0}^{2})^{2}+x^{2}\gamma^{2}},$
\\
where $x=\lambda^{-1}$, $\lambda_{0}$ is the bump peak wavelength, the $c$'s are constant factors for each line of sight and $F(x)$ represents the far UV curvature of the extinction curve. Admittedly, this mathematical representation implies no particular physical model of the ``bump"; however it is of great help in comparing quantitatively our model results with astronomical observations. Since our model carries a negligible amount of underlying continuum (see Fig. 7 to 9), it can only contribute to the term $c_{3}D$. The peak value of this term, at $\lambda_{0}$, is 
\\
$\Delta k(\lambda_{0}-V)=c_{3}/\gamma^{2}$ 
\\
and corresponds to the peak value of $Q/a$. Now, the Lilley empirical relation gives
\\
$E(B-V)=\frac{N_{H}}{5\,10^{21}}$
\\
on average through the ISM, with $N$ the line density of H atoms in cm$^{-2}$. Hence,
\begin{equation}
\Delta E(\lambda_{0}-V)=\Delta k(\lambda_{0}-V)\,E(B-V)\\
=\frac{c_{3}}{\gamma^{2}}\frac{N_{H}}{5\,10^{21}}.
\end{equation}
Also note that, by definition, 
\\
$E(\lambda-V)=1.09\,(\tau(\lambda)-\tau(V))$,
\\
where $\tau$ is the optical depth. So, $\Delta E(\lambda-V)=1.09\,\Delta \tau$, where 
$\Delta \tau$ is the bump in the optical depth.

On the other hand, if $N_{gr}$ is the line density of bump carriers along the line of sight, and if Rayleigh's approximation applies, then the bump in the optical depth is
\\
$\Delta \tau=\pi\,a^{2}\,N_{gr}\,Q$,
\\
assuming the grains are spherical with radius $a$, and there is no line saturation. Also, if $\rho$ is the density of the grains, the line density of carbon atoms locked in the grains is 
\begin{equation}
N_{C}=N_{gr}\frac{4 \pi a^{3} \rho}{3\times 12\times 1.7\,10^{-24}}
\end{equation}
Combining eq. 10 and 11, one finally obtains 
\begin{equation}
\frac{N_{C}}{N_{H}}=20\frac{c_{3}\rho}{\gamma^{2}(Q/a)_{0}}.
\end{equation}

One can see, from the compilation by Fitzpatrick and Massa \cite{fm07}, that the average values of $c_{3}$ and $\gamma$ are, respectively, about 3 and 1 $\mu$m$^{-1}$ (the corresponding bump area is 5$\mu$m$^{-1}$). Take 2 gm/cm$^{3}$ for the density of graphite and, accounting for some disorder in our model grains, assume the average peak value of $Q/a$ is $1.5\,10^{6}$ cm$^{-1}$. Then, on average,  the fraction of cosmic carbon required to be in the model grains turns out to be $\sim80$ ppm, or about 1/4 of the available carbon (for comparison, see Snow and Witt 1995). Of course the largest fraction must be reserved for molecules and for the less ordered and pure carbon grains. As stated earlier, the latter retain their $\sigma$ resonance well beyond the complete damping of their $\pi$ resonance. They may, therefore, well account for the observed rise of interstellar extinction towards shorter wavelengths.

\subsection{Minimum graphene layer size}

 The present model invokes stacks of small graphene sheets. We must, therefore, inquire whether the properties of graphene still apply to sizes smaller than the Rayleigh limit. These properties are determined by the interactions between an atom and its neighbours. Obviously, these interactions fade away as the distance to the central atom increases. In order to find the ``cut-off" distance, we consider compact clusters of benzene rings of increasing size. With the help of a commercial Molecular Modeling package [Hypercube,
Hyperchem 7 \cite{hyp}], their structure was optimized by seeking the minimum total binding energy,  using the Quantum Mechanical semi-empirical AM1/UHF method. Figure 14 shows the variation of the binding energy per atom, $e_{b}$, as a function of the number of rings, N$_{\rm{r}}$.

\begin{figure}
\resizebox{\hsize}{!}{\includegraphics{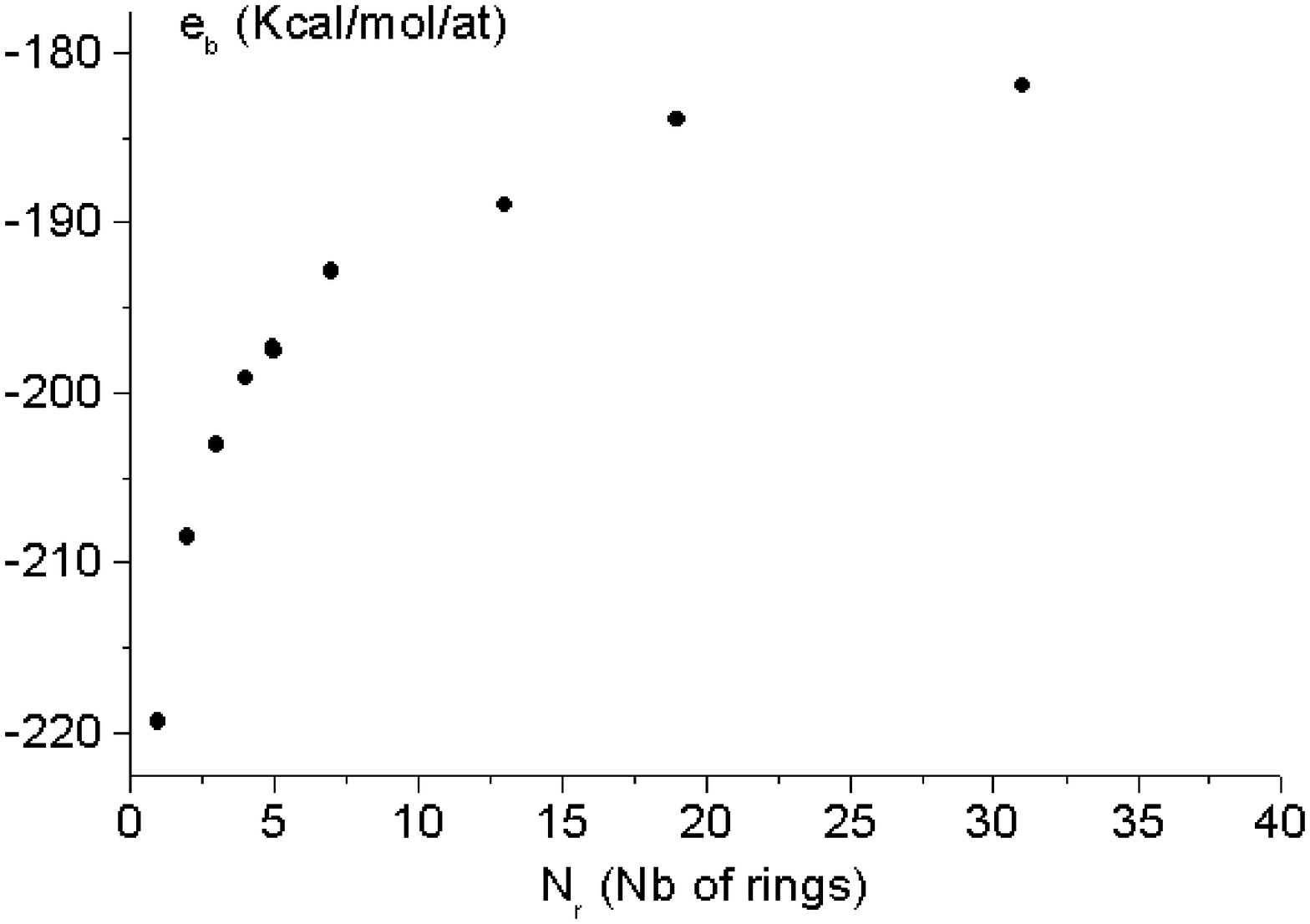}}
\caption[]{Binding energy per atom (Kcal/mol) as a function of the number of benzenic carbon rings in planar, compact clusters;
1 Kcal/mol = 0.0434 eV.}
\label{Fig:figA}
\end{figure}

Beyond about 40 rings, or 84 atoms, the binding energy reaches, for all practical purposes, an asymptotic value of -182 Kcal/mol or 8 eV per atom, which is indeed about the graphite sublimation energy. The corresponding diameter of the structure is about 15~\AA, the size of a large PAH or Platt particle (see Donn 1968). A similar conclusion was reached by Robertson \cite{rob} on the basis of a simpler H\"uckel calculation (see his figures 22 and 28) . Thus the minimum extension required for a graphene sheet is small enough that many such sheets can be accommodated in a Rayleigh-sized grain.

\subsection{Effects of grain shape}

The Fr\"ohlich formula for $Q/a$ includes the ellipsoidal characteristic parameter, $L$, which was set above at 1/3, assuming a spherical shape for both the grain and its subgrain components. As noted by Bohren and Huffman \cite{boh}, shape effects are no weaker in the Rayleigh approximation than for large particles, and the band shape of discs or needles are very different  from the ``Drude" profile. Fortunately, the physics of carbon grain formation and evolution apparently does not allow the evolution towards such extreme shapes. Besides, the grain surface is certainly very irregular and may not be subjected to the same theoretical treatment as perfectly smooth ellipsoids. Finally, even though the term ``surface mode" is usually applied to the Fr\"ohlich resonance, it is clear that the whole grain volume is involved, so the surface shape may not be so important, after all. In order to substantiate these conjectures, we also computed $Q/a$ for values of $L$ symmetrically bracketing 1/3, from 0.25 to 0.4, corresponding to ratios of principal axes ranging between 1 and about 2. It was found that, although the peak wavenumber of the feature increases from 4.4 to 4.82 $\mu$m$^{-1}$, the average feature is still narrow (1 $\mu$m$^{-1}$) and does not shift much away from 4.68 $\mu$m$^{-1}$. This is due to the symmetry of the variations, a consequence of the implicit assumption that the fluctuations of L around 1/3 are limited and symmetric.

\subsection{Turbostratic disorder}

Inter-atomic forces naturally tend to maintain adjacent graphene layers in the preferred stacking schemes, ABA or ABC. Recently, however, slightly disoriented graphene multi-layers have attracted attention in the laboratory (see Kim et al. 2012; Trambly de Laissardi\`{e}re et al. 2010). We also considered the impacts of this type of disorder upon the dielectric function. For this purpose, we performed preliminary calculations, using a simplified tight-binding theory (assuming constant electric dipole moment). The high energy wing of the $\pi$ resonance was found to be quite insensitive to disorientation, and so is the UV bump. But, interestingly, the resonance peak is split into 2 peaks, whose separation increases with the angular deviation ($\sim1$ eV for 10 deg between adjacent layers). This effect could contribute to the flattening of the resonance peak of terrestrial and laboratory graphite (see Taft and Philipp 1965).

\subsection{Hydrogenation}

\begin{figure}
\resizebox{\hsize}{!}{\includegraphics{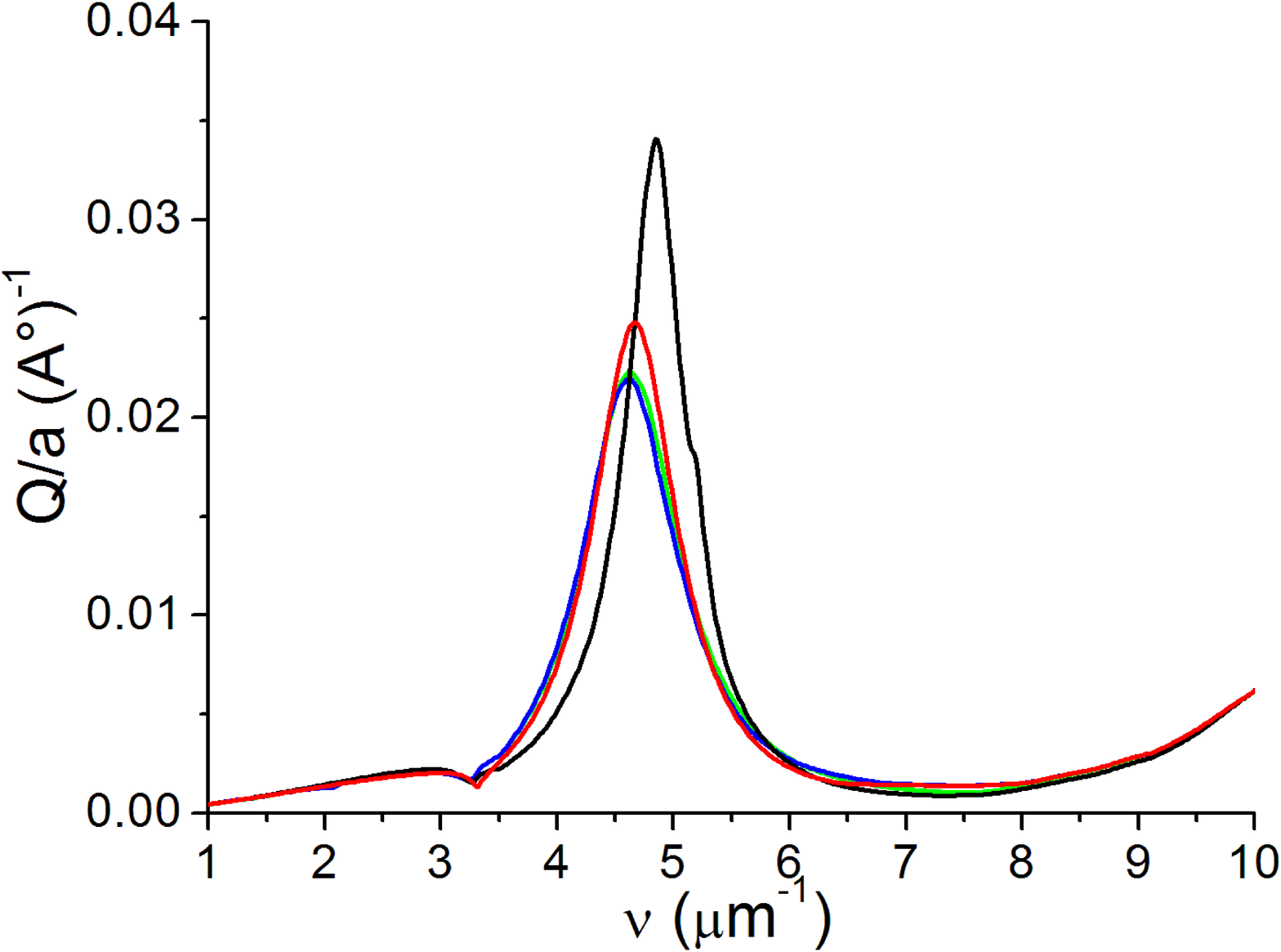}}
\caption[]{Effect on the UV bump of hydrogenation (H/C) of the faces of graphene stacks. $red$: clean faces; $green,\,blue \,and \,black$: H/C(per face)=50, 75 and 100$\%$, respectively. The corresponding widths are 1, 1 and 0.65 $\mu$m$^{-1}$; and the peak wavenumbers: 4.63, 4.61 and 4.84  $\mu$m$^{-1}$.}
\label{Fig:Qhydro}
\end{figure}
 Duley and Seahra \cite{dul98} considered in detail the effects of edge hydrogenation on various regular aromatic structures. They noted some positive correlations between $\gamma$ and hydrogenation (their Fig. 11). For partial hydrogenation, there are also some weak positive correlations between $\nu_{0}$ and hydrogenation. 

In our treatment, we use rectangular graphene sheet with armchair and zigzag edges. The size of a sample with $4096\times 4096$ carbon atoms is about 
$8700\times 5000\ \approx 0.44 \mu$m$^{2}$. This is large enough that its electronic properties are largely insensitive to the (open or periodic) boundary conditions (see Yuan et al. 2010). We assume, here, that this is also true for the flakes we consider in the dust model of Fig. 11 ($\sim100\times100\,\AA^{2}$ in size). It must be kept in mind, of course, that edge hydrogenation of large PAHs was shown to affect notably $\pi-\pi$* transitions (Duley and Seahra 1998; Malloci et al. 2008). However, the maximum size of PAHs considered in these studies was smaller than $7\times7$ rings, which is way smaller than our assumed model dust.

Internal random hydrogenation considered in the present work is totally different from edge hydrogenation in that it introduces $sp^{3}$ hybridization, but it was also found to increase both $\nu$ (slightly) and $\gamma$ (see Fig. \ref{Fig:FMfig16}).

Another possible ``defect" is the hydrogenation of the outer faces of the edge layers of  stacks of graphenes. The effect on the bump is illustrated in Fig. \ref{Fig:Qhydro} for 3 degrees of hydrogenation: H/C=50, 75 and 100 $\%$. It appears that partial hydrogenation only broadens slightly the bump. Complete hydrogenation, however, seems incompatible with observations. This may mean that hydrogenation in space is only partial. But this may also be due to the fact that our tight-binding model for hydrogenated graphene only includes the $\pi$ band, and is therefore not accurate for large concentration of hydrogenation, in which there is a strong overlap between the $\sigma$ and $\pi$  bands. A tight-binding model including both  $\sigma$ and $\pi$ bands for hydrogenated graphene is currently under development.

 By
 contrast, the absence of the $\sigma$ bands in our present treatment is not expected to affect the behavior of graphene upon the introduction of the other two types of disorder considered in this work (random on-site potential, random hopping constant); for the latter do not add any extra terms to the Hamiltonian.

 \subsection{Ionization}
Ionization of the grains is likely to occur in the ISM and should therefore be considered here. We found that an electronic charge in the order of ~0.1 electron per C atom will only shift the Fermi level by ~0.2 eV, leaving the spectrum at higher energies unchanged.  
In fact, taking into account that the $\pi$ resonance is associated to inter-band transitions from the Van Hove singularity of the valence band to the Van Hove singularity of the conduction band, both of them being separated by an energy window larger than 4eV, weak amounts of doping will not modify the optical spectrum enough to affect the main features of the resonant peak of the $\pi$ mode.(see Yuan et al. 2011a).

On the other hand, Cecchi-Pestellini et al. \cite{cec} showed that, for several PAH's, the spectra of neutrals, anions, cations and dications were different between 6 and 10 eV. This effect may be due to interaction between the $\pi$ and $\sigma$ resonances. Addressing this issue will become possible when both resonances are included in the tight-binding model. 

 \subsection{Work in progress}
Clearly, a valuable addition to this work would be the inclusion of the $\sigma$ resonance in the calculation of the DOS and the optical conductivity. Work is in progress to fill this gap. To include both $\sigma$ and $\pi$ bands one needs to consider four orbitals (s, px, py and pz) on each carbon atom and one orbital (s) on each hydrogen atom. The dimension of the hopping matrix between two nearest carbon atoms becomes $8\times 8$ , which makes the computation more expensive as compared to the single $\pi$ band model which only includes a $2\times 2$ hopping matrix. Each $8\times 8$ hopping matrix between two neighbours has to be recalculated according to the coordinates of each carbon atom. To the authors' knowledge the full tight-binding model including both $\sigma$ and $\pi$ bands for graphene is not available in the literature yet.

At the same time, we are investigating whether the accuracy of the present tight-binding calculation can be improved, for instance, by using larger samples to have more random complex coefficients in the initial wave function, or using more time-steps to increase the energy resolution in the spectrum. The fundamental operation in our numerical methods is the time-evolution of the wave function by using the Chebyshev polynomial method, which has already the same accuracy as the precision of the simulation machine.

\section{conclusion}

The present model carrier consists of nano-sized ``bricks'', i.e. stacks of a few finite sheets of graphene. The physics of carbon is now known in such detail that it has become possible to compute the dielectric functions of such particles from first principles, with no $ad \,hoc$ assumptions, nor tailoring of any sort. In its clean and ordered form, this structure carries a UV bump at 4.65 $\mu$m$^{-1}$, with FWHM 0.85 $\mu$m$^{-1}$, very near the lower limit set by observations. This width is essentially determined by the high-energy slope of $\epsilon_{2}$ as a function of frequency, which, in turn, depends heavily on the variation of the electric dipole moment with frequency. Unlike theories which assume for simplicity that the dipole moment is constant, the present treatment explicitely includes its variations. Because the dipole moment decreases notably as frequency increases, the slope of $\epsilon_{2}$ becomes notably steeper, and the bump width distinctly narrower.

Most natural or artificial carbonaceous materials are neither pure nor perfectly regular in structure. Impure or disordered structures can also be simulated theoretically by allowing the natural energetic parameters of pure graphene to vary according to a statistical distribution, thus simulating different types of structural, chemical and orientation disorder. We have found that three types of structural/chemical disorder affect notably the high-energy slope and, hence, broaden the UV bump. Moderate degrees of disorder are enough to cover the observed range of widths while the peak position remains constant to $\sim1\,\%$.

The peak position is more sensitive to the type of disorder than to its degree. For any of these types of disorder, the general trend is for the width to increase with the degree of disorder. Orientation disorder of bricks randomly glued together in one dust particle, brings about more broadening and peak shift. This dust model, too, is sensitive to the structural and chemical disorders considered in this paper, with similar consequences on position and breadth of the peak. When all cases are lumped together in the same diagram, no obvious correlation emerges between bump width and position.

In essence, our theoretical procedure boils down to replacing $sp^{2}$ by $sp^{3}$ or $sp^{1}$ bonds. No wonder, then, that this leads, in particular, to a depression of the $\pi$ resonance, which is observed in various amorphous carbons and hydrocarbons or coals of various ages. Together with the general trend of increasing width and decreasing strength of the resonance, the results that we present here can help to draw  parallels, and identify common features, between the different proposed bump-carrier models.

\section{Acknowledgments}
We are indebted to the referee, Dr G. Mulas, for comments which helped to improve this paper. 
S.Y., R.R. and M.I.K. acknowledge a financial support of the EU-India FP-7 collaboration
under MONAMI. S.Y. and M.I.K. acknowledge the  Netherlands National Computing Facilities foundation (NCF) for the use of supercomputing facilities. R. R. also acknowledges financial support from the Juan de la Cierva Program (MINECO, Spain).

\end{document}